\documentclass[12pt]{article}

\pdfoutput=1 
\newcommand{\frb}{FRB\,20190520B}
\usepackage{longtable}
\newcommand{\dmunits}{pc cm$^{-3}$}
\newcommand{\rmunits}{rad m$^{-2}$}
\usepackage{color}

\usepackage{graphicx}
\usepackage{url}
\usepackage{pdflscape}
\usepackage{scicite}
\usepackage{hyperref}
\usepackage{authblk}
\usepackage{times}

\newenvironment{sciabstract}{%
\begin{quote} \bf}
{\end{quote}}

\title{Magnetic field reversal in the turbulent environment around a repeating fast radio burst}

\author
{
Reshma Anna-Thomas,$^{1,2\ast\dagger}$
Liam Connor,$^{3,4\dagger}$ Shi Dai,$^{5,6,7\dagger}
$Yi Feng,$^{8\dagger}$ Sarah Burke-Spolaor,$^{1,2\dagger}$ 
Paz Beniamini,$^{9,10}$ 
Yuan-Pei Yang,$^{11}$Yongkun Zhang,$^{6}$
Kshitij Aggarwal,$^{1,2}$ 
Casey J. Law,$^{3,4}$ 
Di Li,$^{6,8,12,13\ast}$ Chenhui Niu,$^{6}$
Shami Chatterjee,$^{14,15}$ Marilyn Cruces,$^{16}$ Ran Duan,$^{6}$ 
Miroslav D. Filipovi,$^{5}$ George Hobbs,$^{7}$ Ryan S. Lynch,$^{17}$ Chenchen Miao,$^{6}$ Jiarui Niu,$^{6}$
Stella K. Ocker,$^{14,15}$ Chao-Wei Tsai,$^{6,12,22}$ Pei Wang,$^{6}$
Mengyao Xue,$^{6}$  Jumei Yao,$^{18}$ Wenfei Yu,$^{19}$ Bing Zhang,$^{20,21}$ Lei Zhang,$^{6}$ Shiqiang Zhu,$^{8}$ Weiwei Zhu$^{6,22}$
\\
\footnotesize{$^{1}$Department of Physics and Astronomy, West Virginia University, Morgantown, WV 26506, USA}\\
\footnotesize{$^{2}$ Center for Gravitational Waves and Cosmology, West Virginia University, Morgantown, WV, USA}\\
\footnotesize{$^{3}$Cahill Center for Astronomy and Astrophysics, California Institute of Technology, }
\footnotesize{Pasadena, CA 91125, USA}\\
\footnotesize{$^{4}$Owens Valley Radio Observatory, California Institute of Technology,}
\footnotesize{Big Pine, CA, 93513, USA}\\
\footnotesize{$^{5}$School of Science, Western Sydney University, Locked Bag 1797, Penrith, NSW 2751, Australia}\\
\footnotesize{$^{6}$National Astronomical Observatories, Chinese Academy of Sciences, Beijing 100101, China}\\
\footnotesize{$^{7}$The Commonwealth Scientific and Industrial Research Organisation Space and Astronomy,}\\
\footnotesize{Australia Telescope National Facility, Epping, NSW 1710, Australia}\\
\footnotesize{$^{8}$Zhejiang Lab, Hangzhou, Zhejiang 311121, China}\\
\footnotesize{$^{9}$Department of Natural Sciences, Open University of Israel, Ra’anana 43107, Israel}\\
\footnotesize{$^{10}$Astrophysics Research Center of the Open University, The Open University of Israel,}\\
\footnotesize{Ra’anana 43537, Israel}\\
\footnotesize{$^{11}$South-Western Institute for Astronomy Research, Yunnan University, Kunming 650500, Yunnan, China}  \\
\footnotesize{$^{12}$University of Chinese Academy of Sciences, Beijing 100049, China}  \\
\footnotesize{$^{13}$National Astronomical Observatories, Chinese Academy of Sciences-University of KwaZulu-Natal}\\
\footnotesize{Computational Astrophysics Centre, University of KwaZulu-Natal, Durban 4000, South Africa}\\
\footnotesize{$^{14}$Department of Astronomy, Cornell University, Ithaca, New York 14853, USA}\\
\footnotesize{$^{15}$Cornell Center for Astrophysics and Planetary Science, Cornell University, Ithaca, New York 14853, USA} \\
\footnotesize{$^{16}$Max-Planck Institute for Radio Astronomy, Auf dem Hügel 69, D-53121 Bonn, Germany} \\
\footnotesize{$^{17}$Green Bank Observatory, Green Bank, WV 24401, USA}\\
\footnotesize{$^{18}$Xinjiang Astronomical Observatory, Chinese Academy of Sciences, Urumqi, Xinjiang 830011, China}\\
\footnotesize{$^{19}$Shanghai Astronomical Observatory, Chinese Academy of Sciences,}
\footnotesize{80 Nandan Road, Shanghai 200030, China} \\
\footnotesize{$^{20}$Nevada Center for Astrophysics, University of Nevada, Las Vegas, NV 89154, USA}\\
\footnotesize{$^{21}$Department of Physics and Astronomy, University of Nevada, Las Vegas, NV 89154, USA}\\
\footnotesize{$^{22}$Institute for Frontiers in Astronomy and Astrophysics, Beijing Normal University,  Beijing 102206, China} \\

\footnotesize{$^\ast$Corresponding author; E-mail:  rat0022@mix.wvu.edu, dili@nao.cas.cn}\\
\footnotesize{$^\dagger$These authors contributed equally to this work.}
}

\date{}

\begin{document} 


\baselineskip24pt


\maketitle

%
%
%
%


\begin{sciabstract}
Fast radio bursts (FRBs) are brief, intense flashes of radio waves from unidentified extragalactic sources. 
Polarized FRBs originate in highly magnetized environments. We report observations of the repeating \frb\ spanning seventeen months , which show its amount of Faraday rotation is highly variable and twice changes its sign. The FRB also depolarizes below radio frequencies around 1$-$3 GHz. We interpret these properties as due to change in the parallel component of the integrated magnetic field along the line-of-sight, including reversals. This could result from propagation through a turbulent, magnetized screen of plasma located between $10^{-5}$ to $100$ parsecs of the FRB source. This is consistent with the bursts passing through the stellar wind of a binary companion of the FRB source.

\end{sciabstract}


Fast radio bursts (FRBs) are millisecond flashes of radio waves \cite{Lorimer2007} from distant galaxies. Their emission mechanism, sources and local environments are not well understood. The magnetization and density of astrophysical plasmas along the line-of-sight (LOS) between an FRB source and Earth modify the FRB's polarization properties and can provide constraints on the source's local environment. FRB discoveries exhibiting new properties often have provided insight into the media and origins of these enigmatic sources, for instance observations of highly magnetized and variable plasma environments in FRBs \cite{MICHILLI2018,Xu2022,McKinven2022_180916RM}, detection of FRB-like bursts from a known Galactic magnetar \cite{chime1935,Bochenek20201935} and detection of `nano-shots' in FRB bursts comparable to Crab pulsar \cite{Nimmo2021+20200120E}.

The repeating \frb\ has been localized to a dwarf galaxy at redshift $z=0.241\pm 0.001$ \cite{Niu2021}. \frb\ has a sustained high repetition rate at 1.2 GHz above a given fluence and is given by  $R_{\rm 1.2GHz}(>9.3{\rm \,mJy\,ms}) = 4.5^{1.9}_{-1.5} {\rm hr}^{-1}$ (where $ \rm 1 mJy = 10^{-29} W\,m^{-2}\,Hz^{-1}$). It is co-located with a compact persistent radio source (PRS), and has a large contribution from the host galaxy to its LOS electron column density. The integrated electron column density $n_e(l)$ along the LOS distance, $l$, is quantified by the dispersion measure, ${\rm DM} \equiv \int n_{\rm e}(l) dl$. For \frb, the DM is larger than the foreground Milky Way and the intergalactic medium (IGM), which dominate the DMs of most other FRBs \cite{Macquart2020}. The rest-frame contribution of the host galaxy to the DM of \frb  is $\rm DM_{host} = 902^{+88}_{-128}$ \dmunits \cite{Niu2021,ocker+22}, (where $\rm 1~pc =3.08\times10^{16}~m$) about an order of magnitude larger what is typical for FRB hosts, implying a large number of free electrons in the local environment \cite{Macquart}. \frb\ has several similarities with the repeating FRB\,20121102A (05h31m59s, +33d08$'$53$''$) \cite{Chatterjee2017,Tendulkar2017} - both are located in dwarf galaxies with low abundance of heavy elements and are associated with a PRS. Another important measure, the Faraday rotation measure (RM), quantifies the amount of Faraday rotation of the radio waves integrated along the LOS distance,$l$, given by $RM = [e^3/(2\pi m_{\rm e}^2 c^4)]\int n_e(l)B_{||}(l)dl$, 
where $B_{||}$ is the component of the magnetic field parallel to the LOS, $e$ is the charge of an electron, $c$ is the speed of light and $m_{e}$ is the mass of  an electron. Faraday rotation is termed as the rotation of the plane of polarization of a linearly polarized wave when it propagates through a magnetized plasma \cite{FaradayRotation1846}. FRB\,20121102A has a high RM \cite{MICHILLI2018}, but the RM of \frb\ has not previously been monitored. 

\section*{Bursts observed from \frb}
We conducted polarimetric observations of \frb\ using the 100-m Robert C. Byrd Green Bank Telescope (GBT) and the 64-m Parkes Radio Telescope (also known as Murriyang). The GBT observations used the 1.1 to 1.8\,GHz (L-band) receiver on Universal Time Coordinated (UTC) 2020 September 17, and with the 4 to 8\,GHz (C-band) receiver on 2020 September 14 (epoch 1), 2021 March 19, 23, 27 and 31 (epoch 2), and 2021 November 4 (epoch 3). The observations totaled 6.5 hours at L-band and $\sim$\,20 hours at C-band \cite{methods}. The Parkes observations were performed fortnightly from  2021 April 5 to 2022 January 23 using the Ultra-Wideband Low (UWL) receiver, which covers 704 to 4032\,MHz \cite{2020PASA...37...12H}. We performed a threshold-based search for bursts in these data, including a search of frequency sub-bands to find spectrally narrow-banded bursts \cite{methods}. In total, we found 9 bursts in the GBT L-band data, 16 bursts in the GBT C-band data, and 113 bursts in the Parkes data (between $\sim1600$ and 4032\,MHz) with each having a signal-to-noise ratio (S/N) $\ge$ 10. We performed polarimetric and flux calibration on each detected burst data segment \cite{methods}.

The burst profiles and spectrograms of 12 bursts are shown in Fig.~S2, and the properties of all the thirteen bursts with polarization detections are listed in Table~S2; DM and polarization upper limits for all other bursts are reported in Table~S3. The Parkes bursts with polarization detection are named P1 to P8 sorted by their Modified Julian dates (MJDs), and the GBT C-Band bursts with polarization detection are named C1, C2.1, C2.2, C3.1 and C3.2, where the integral part represent the epoch of the GBT observation it was detected and the fractional part represent the order of the arrival time of the burst.
The detected bursts are spectrally narrow-banded, as has been seen for other repeating FRBs \cite{Kumar2021,Gajjar2018, Pleunis2021Chime}, and some pulses have time-resolved sub-structure. The average detected burst bandwidths are 350 MHz in the GBT L-band data, 850 MHz in GBT C-band, and 750 MHz in the Parkes UWL data.

We analyzed the Stokes paramters to determine the linear polarization fractions and RMs of the bursts. We find that the polarization of \frb is time- and frequency-dependent, and the RM is also variable (Fig.~1). The RM detections, found by maximizing the linear polarization fraction are shown in Fig.~2. After removing the Faraday rotation, 
5 GBT C-band bursts and 8 UWL bursts have fractional linear polarization $(L/I)$ (where $L$ is the linear polarization and $I$ is the total intensity), in the range 15$\%$ to 80$\%$. Only one burst (C2.1) had detectable fractional circular polarization, with $V/I$ (where $V$ is the circular polarization) of $-42\pm7$\%. No L-band bursts had detectable polarization, with the brightest burst having an upper limit of $L/I\leq9\%$. We estimate that the L-band data had sufficient frequency resolution to recover RMs up to $10^{5}$ \rmunits~, so this non-detection is not due to instrumental depolarization \cite{methods}. The spectrogram and the profile of three bursts that demonstrate depolarization towards low frequency is shown in Fig.~3.
The low linear polarization in the L-band is an intrinsic property of the FRB, as is the variable polarization in the C-band and across the UWL band. They could be due to the emission properties of the source, propagation through an intervening plasma screen, or a combination of the two effects.
 
For the thirteen bursts with detected polarization, we measured RMs ranging from  $-2.4 \times 10^{4}$\,rad\,m$^{-2}$ in the observer's frame (corresponding to $-3.6 \times 10^{4}$\,rad\,m$^{-2}$ in the source frame) to $+ 1.3 \times 10^{4}$\,rad\,m$^{-2}$ in the observer's frame (corresponding to $+2.0 \times 10^{4}$\,rad\,m$^{-2}$ in the source frame). The foreground Milky Way contribution to the RM in this direction is small, $\rm -9.85 \pm 12.64$ \rmunits \cite{Opperman2012}. We observed RM variability on week-long timescales of around $\pm$300\,rad\,m$^{-2}$\,day$^{-1}$, and similar average variability on six-month timescales. The largest inter-epoch rate of RM change, $dRM/dt=-444\,$\rmunits\,day$^{-1}$, between 2021 June 8th and 2021 June 19th. 

\section*{Comparison with other sources}
The range of RM we measure for \frb\ spans---in a single target---the full range of negative and positive RMs observed for pulsars in the Milky Way (Fig.~4). 
While the maximum absolute RM value for FRB~20190520B is three times lower than that of FRB~20121102A, their peak-to-peak variations are of the similar size. Therefore, the fractional change in RM for \frb\ is much larger. Adding the fact that the RM of FRB~20190520B crosses zero instead of simply varying, this implies a magnetized plasma environment of similar properties, however requiring an additional environmental consideration to allow sign reversals.
The frequency-dependent polarization fraction change for \frb\ indicates multi-path depolarizing effects, while the fractional RM change and the sign reversals require changes in the LOS magnetic field (either due to rapid reorientation of a bulk material or integrated changes due to turbulence).

The Galactic source with the most similar RM variability to \frb\ is PSR B1259--63 (13h02m48s, -63d50$'$09$''$, Fig.~4), a pulsar that is periodically eclipsed by its binary star companion \cite{bestar}. Both \frb\ and PSR B1259--63 have RM variations that cross zero. This indicates a reversal in the integrated LOS magnetic field orientation, in addition to variation in the electron density or magnetic field strength.
The time-dependent variations of RM in both sign and magnitude in PSR B1259--63 have been attributed to passage of the radio pulses through a clumpy decretion disk around its binary star companion during the closest part of the orbit, causing depolarization \cite{Johnston1996}. The pulsar's DM rose by around 20\,\dmunits\ over a few days just prior to the eclipse part of its orbit. Other FRBs, pulsars and a magnetar which shows RM variations have been discussed in the Supplementary text.

\section*{Dispersion Measure}
Whatever the source of the RM variations, if $\int_{l} n_e\,dl$ along the LOS contributes to the RM magnitude changes for \frb, there should be an accompanying DM change. We therefore seek to quantify DM changes, although this is complicated by the sad trombone effect \cite{Hessels2019} that is exhibited by some FRBs. The sad trombone effect is the observation of pulse structures that sweep downwards in frequency as a function of time, as is visible for some bursts of \frb\ in Fig.~S3. We therefore inspected the brightest pulses (detection S/N $>20$) using a common DM measurement technique that maximizes the structure in the burst \cite{methods}, in addition to visual inspection of pulses with substructures. We measure a $\Delta\rm DM$ range of 10 to 15~\dmunits, although the associated uncertainties prevent us from determining whether these vary in a systematic way \cite{methods}. Assuming the changes in DM and RM arise from the same medium, 
we calculate the average LOS magnetic field to be $B_{||}\simeq 1.2 \left( \frac{\rm \Delta RM/(10^4 rad~m^{-2})}{\rm \Delta DM/(10~pc~cm^{-3})}\right)~{\rm mG} \simeq 3$ to 6\, mG. This calculation assumes a fully turbulent medium, for which the $\rm \Delta DM$ of a plasma screen is approximately equal to the average DM contributed by that screen. For comparison, for FRB~20121102A the lower limit range on the average magnetic field is 0.6 to 2.4 mG \cite{MICHILLI2018}.


\section*{Interpretation of the magnetized environment}
We interpret \frb's properties as due to propagation of the radio waves through a dense, turbulent magnetized region, e.\,g.\ as discussed in \cite{2021arXiv211000028B}. We do not consider large-scale magnetic reorientations of a LOS object, or something arising in a more diffuse origin in the host (e.g. time-variable LOS due to a spiral arm), as viable descriptors due to the large variations and rapid timescales involved. We also do not explore here explanations intrinsic to the FRB emission mechanism; while this is possible, the observed properties seem to be well-fit to a propagation model, as explored here and in the supplementary material.
In our preferred scenario, the polarization, RM and DM are imparted in a region dominated by bulk magnetization, where the region is made up of sub-eddies or filamentary regions which cause time variability in the integrated LOS magnetic field or electron density

 Multi-path depolarization and large RM variability can be produced by turbulent dynamic and dense magnetoionic plasma environments\cite{2021arXiv211000028B}. However, the magnitude of those effects---and therefore whether they can be detected in observations--- depends on the physical parameters of such a screen. 
The observable effects are determined by the characteristic size of turbulent eddy regions, the screen depth, and the relative velocities of the FRB source, bulk screen, and internal eddies. Using a basic model for propagation through such a turbulent screen \cite{methods}, our measurements of \frb constrain the screen to be between $10^{-5}$ pc and $100$\,pc from the FRB source with a free electron density  between $5\times10^6$\,cm$^{-3}$ and 2\,cm$^{-3}$, respectively; note that these results will cover a broader range if the LOS DM variation is substantially lower \cite{methods}. A previous model of FRB 20121102A \cite{MICHILLI2018} found that for a plasma with electron temperature $T_{e}$ = $10^{6}$ K, the electron density $n_{e}= 10^{2}~\rm cm^{-3}$ and the thickness of the plasma screen $L_{RM} = 1~\rm pc$, which is similar to the upper limit on the size of the PRS. Higher electron temperatures could increase the thickness to $L_{RM} = 100~\rm pc$ \cite{MICHILLI2018}.

We considered several scenarios which could explain such environmental turbulence local to an FRB source. Informed by the similar properties of PSR B1259--63, we explore a binary model in which the LOS to \frb\ passes close to the surface of a companion star, so the radio propagation is affected by the magnetization and turbulence of its stellar wind. In this scenario, the star provides the bulk magnetization that causes the broad range in RM, while the turbulent stellar wind causes the rapid RM swings \cite{methods}.  The distance to the plasma screen is then approximately the separation between the two objects. Taking $d\approx 10^{-5}$\,pc, the required electron density is $n_{e}\approx 5\times 10^6\mbox{ cm}^{-3}$, which could be provided by a mass loss rate $\dot{M}\approx 3\times 10^{-8}M_{\odot} \mbox{ yr}^{-1}$ (where $M_{\odot}=2\times10^{30}$ kg) and wind velocity $v_{\rm w}\sim 10^3\mbox{km s}^{-1}$ \cite{methods}. These values are consistent with values observed for massive stars, e.g. \cite{langer-masslossrate}.  We applied the same model to the PSR B1259--63 system (see Supplementary Text), deriving properties that are consistent with previous studies \cite{bestar}. 

If the binary wind scenario is correct, we expect there to be an underlying periodicity to the observed properties (a rise and fall of the RM variability envelope, DM variability range, and average DM) equal to the orbital period. During times when the LOS to the FRB source does not pass close to the star, we expect the RM value to stop varying and settle at the ambient value that reflects the bulk magnetization of the host galaxy. This effect has been observed for pulsars B1259--63 and B1744--24A (17h48m02s, -24d46$'$37$''$)\cite{Johnston2005,Dongzi2022}, 
and other FRBs with RM variability \cite{Xu2022,McKinven2022_180916RM}.

In addition to the above discussion, we considered models of a shocked neutron star wind, or an FRB source close to an intermediate mass black hole, but regard both scenarios as unlikely (see Supplementary Text).

\section*{Implications for other sources}
Although the RM variations and host DM of \frb\ are unsual, they do not necessarily indicate it has an unusual FRB source. \frb\ appears has a pulse rate and spectrally narrow-banded pulse sub-structures that vary with time, consistent with other repeating FRBs, and has similar energy scales to other FRBs \cite{Niu2021,li2021}. It is possible that repeating FRBs have a common source type, but vary in local conditions (e.g. binary orbital period, eccentricity, phase, or inclination). 

Previous studies have proposed a connection between the emission cycles seen in some repeating FRBs and a binary orbit\cite{Barkov2022,Ioka+Zhang(2020),Tendulkar2021}. If our model is correct, the observed properties of the FRB constrain the binary configuration and some properties ($n_e$, magnetic field strength) of the stellar wind.

\clearpage
 \begin{figure*}
     \centering
     \includegraphics[width=0.84\textwidth]{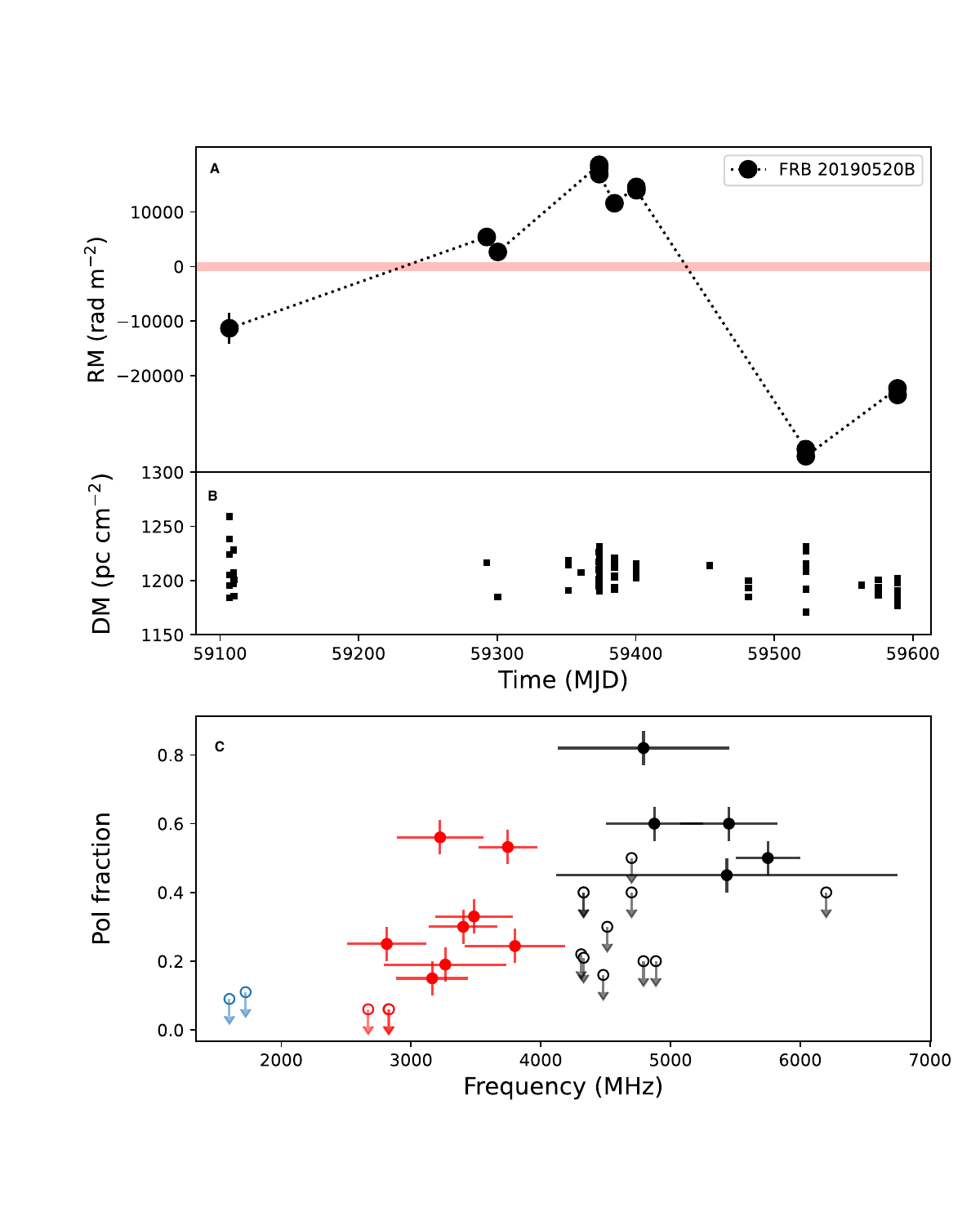}
     \caption{\textbf{Observed time- and frequency-dependent properties of \frb}. (A) show the rest-frame RM. The pink shaded region denotes 0 \rmunits. (B) shows DM variation over a 17-month period. $1\sigma$ uncertainies on RM and DM are also plotted. (C) shows polarization fraction as a function of observed frequency, with data points indicating the GBT L-band (blue), C-band (black) and Parkes data (red). Solid symbols are detections and empty symbols with downward pointing arrows are 3$\sigma$ upper limits. Only the most constraining upper limits is shown at each frequency to avoid overcrowding.  Data for all bursts are listed in Table.~S3.} 
     \label{fig:rmdm}
 \end{figure*}

\clearpage
 \begin{figure*}
      \centering
      \includegraphics[width=0.7\textwidth]{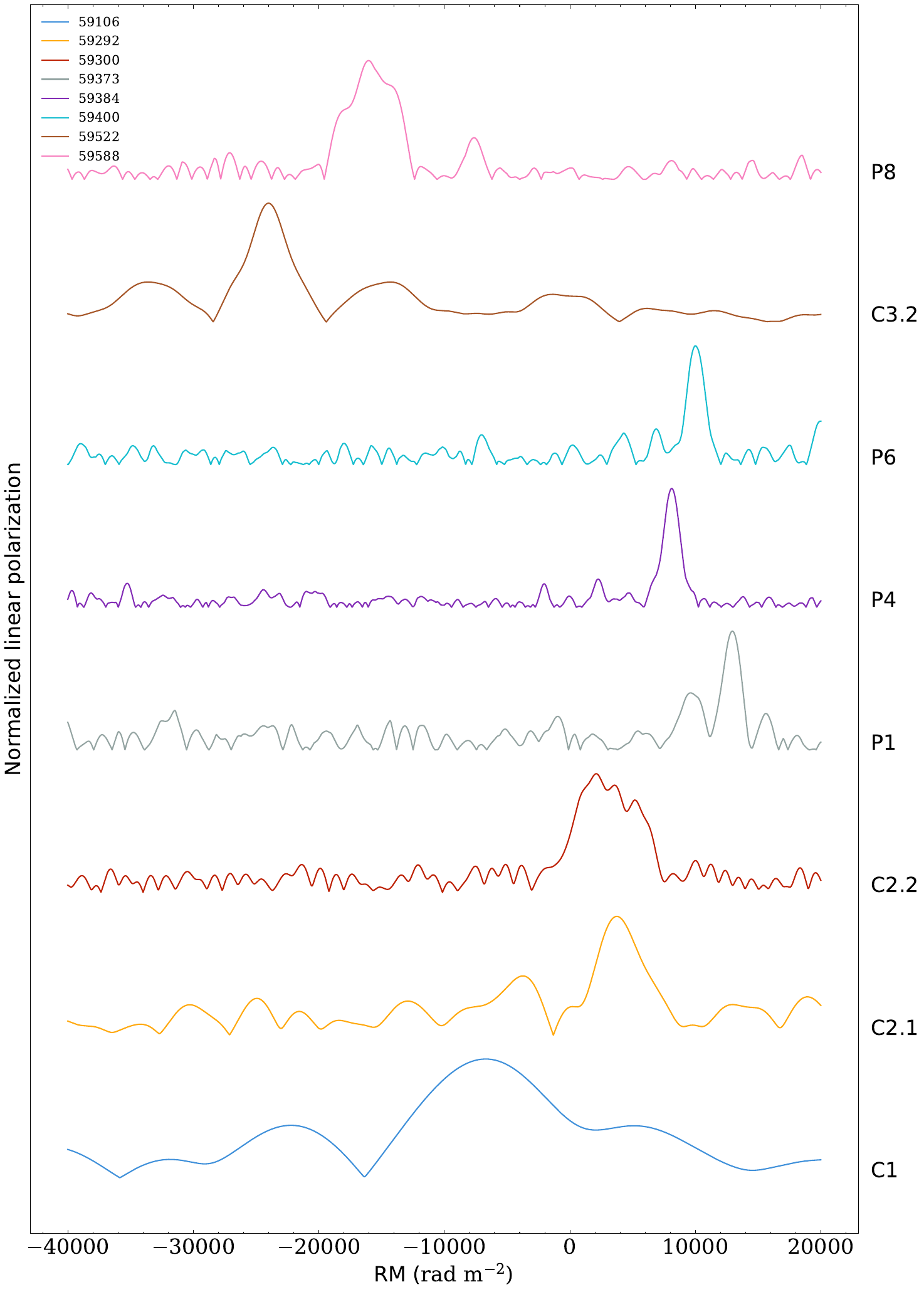}
      \caption{\textbf{Time evolution of RM demonstrated by \frb.} Normalized linear polarization fraction of eight bursts is shown a function of RM. The maximum value of linear polarization determines the RM at the time of each burst. Each trace shows the measurement from a different burst, sorted by date, with arbitrary offsets. The colors represent different MJDs. Only one burst per MJD is plotted here. The burst notations on the right hand y-axis is explained in the Main text.}
      
      \label{fig:faradyspectrum}
 \end{figure*}

\clearpage
\begin{figure*}
    \centering
    \includegraphics[width=1.0\textwidth]{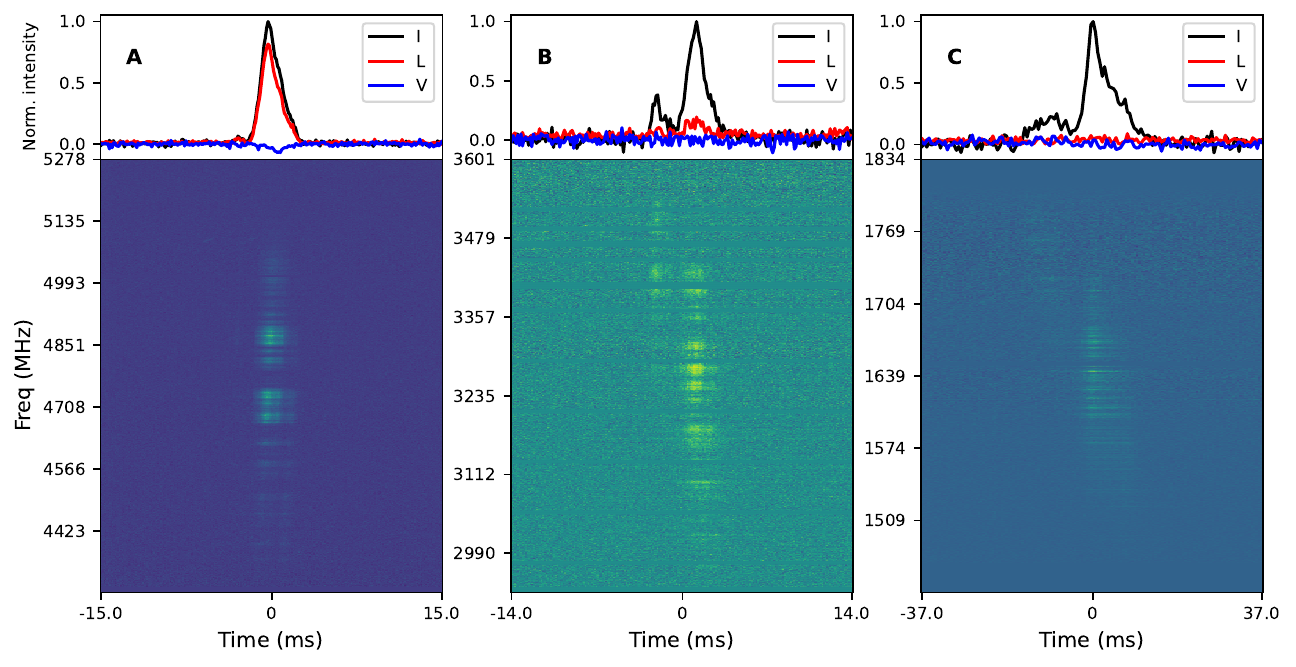}
    \caption{\textbf{Depolarization towards low-frquencies.} The figure shows the spectrogram and the frequency averaged polarization profile of three bursts at different frequencies. The burst in panel A is a GBT C-Band (4.8 GHz) burst and shows the highest percentage of linear polarization in the sample, B is a Parkes UWL (3.2 GHz) burst which shows a small fraction of linear polarization and C is a  GBT L-Band (1.6 GHz) burst and shows no detectable linear polarization. The black, red and blue traces represents the Stokes I, linear polarization L and Stokes V profiles respectively.}
\end{figure*}

\clearpage
 \begin{figure*}
     \centering
     \includegraphics[width=0.85\textwidth]{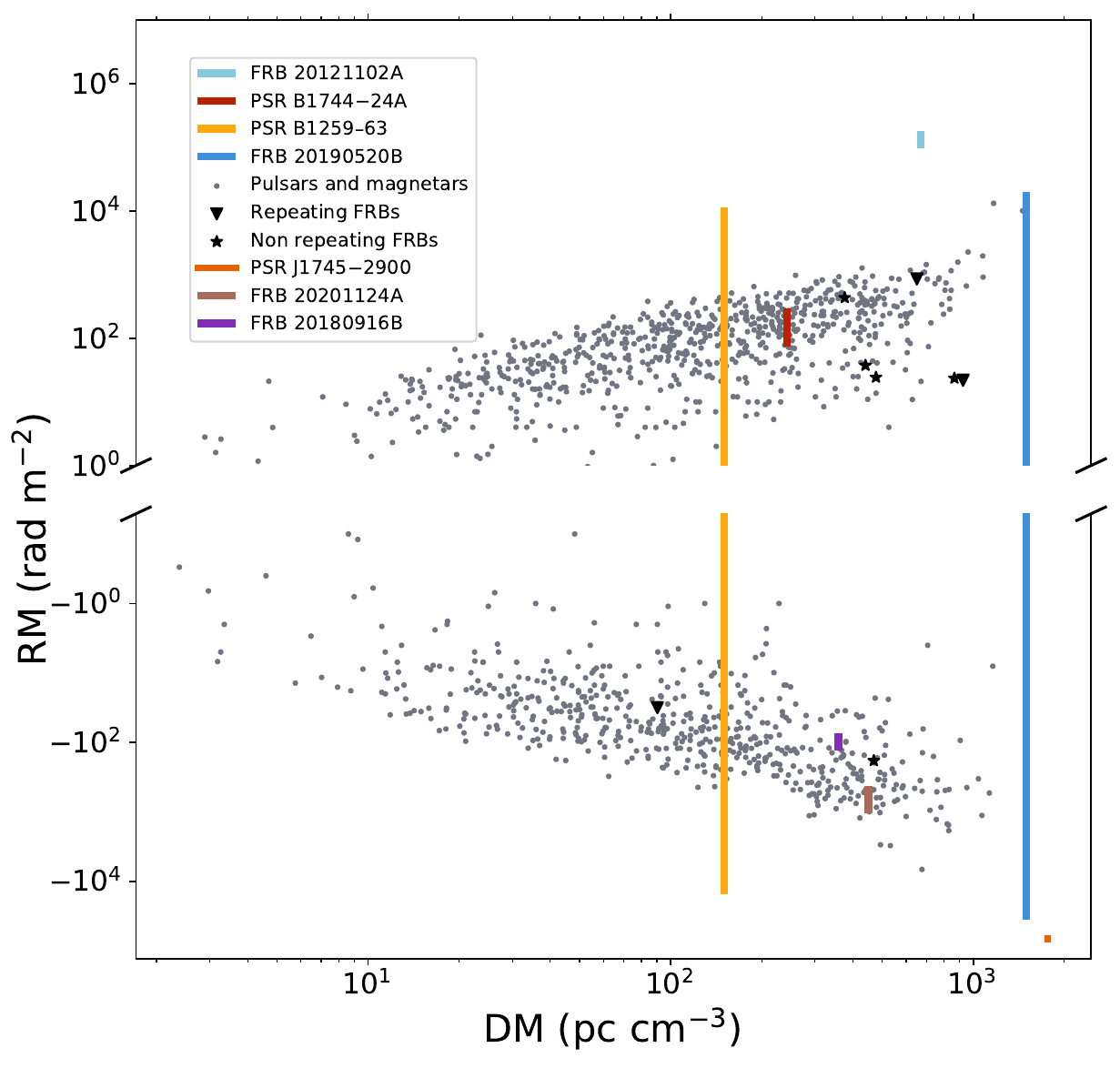}
    
     \caption{\textbf{RM and DM distributions of pulsars, magnetars and FRBs}. The thick vertical lines correspond to the sources that shows RM variations and extends across the range of RM observed in those sources. The grey circles represents pulsars and magnetars\cite{ATNF_psrcat}, the star and downward triangle represents non-repeating and repeating FRBs\cite{frbcat,chime_cat} which doesn't show any RM variations, respectively. \frb\ and PSR B1259--63 have larger fractional variability than other FRBs or pulsars, and both exhibit zero-crossings in RM. As per IAU guidelines we provide approximate positions here for the specific sources mentioned in the format: sourcename (J2000 RA, J2000 Dec): FRB20121102A (05h31m59s, 33d08$'$53$''$) \cite{Plavin2022}; PSR B1744--24A (17h48m02s, -24d46$'$37$''$) \cite{Dongzi2022}; PSR B1259--63 (13h02m48, -63d50$'$09$''$) \cite{Johnston1996,Connors2000,Johnston2005}; FRB 20190520B (16h02m04s, -11d17$'$17$''$); PSR J1745--2900 (17h45m40s, -29d00$'$30$''$) \cite{desvignes+18}; FRB 20201124A (05h08m3s, 26d03$'$38$''$) \cite{Xu2022}; FRB20180916B (01h58m01s, 65d43$'$00$''$) \cite{McKinven2022_180916RM}}.
 \end{figure*}

\clearpage

\clearpage
\bibliography{frb20190520B_pol}

\bibliographystyle{Science}

\section*{Acknowledgments}
The Parkes (Murriyang) Radio Telescope is part of the Australia Telescope National Facility, which is funded by the Commonwealth of Australia for operation as a National Facility managed by CSIRO. The Green Bank Observatory is a facility of the National Science Foundation operated under cooperative agreement by Associated Universities, Inc. 

\section*{Funding}
 DL and YF are supported by NSFC grant No. 11988101, 12203045, 11725313, by the National Key R$\&$D Program of China No. 2017YFA0402600, and by Key Research Project of Zhejiang Lab No. 2021PE0AC03 RAT, SBS, and KA acknowledge support from NSF grant AAG-1714897. SD is the recipient of an Australian Research Council Discovery Early Career Award (DE210101738) funded by the Australian Government. YPY is supported by NSFC grant No. 12003028. CJL acknowledges support from NSF grant No.~2022546. PB was supported by a grant (no. 2020747) from the United States-Israel Binational Science Foundation (BSF), Jerusalem, Israel. SBS is a CIFAR Azrieli Global Scholar in the Gravity and the Extreme Universe program. WWZ is supported by National SKA Program of China No.2020SKA0120200 and the NSFC 12041303, 11873067. PW is supported by NSFC grant No. U2031117, the Youth Innovation Promotion Association CAS (id.~2021055), CAS Project for Young Scientists in Basic Reasearch (grant YSBR-006) and the Cultivation Project for FAST Scientific Payoff and Research Achievement of CAMS-CAS. JMY is supported by NSFC grant No. 11903049, and the Cultivation Project for FAST Scientific Payoff and Research Achievement of CAMS-CAS. LZ is supported by ACAMAR Postdoctoral Fellowship and the National Natural Science Foundation of China (Grant No. 12103069).
MC, DL and WWZ acknowledge support from the CAS-MPG LEGACY project. CWT is supported by NSFC grant No. 12041302. WY is supported by NSFC grant No.U1838203. The GBT Epoch 2 observations were carried out through a time procurement agreement funded by a NSFC grant 11988101.

\section*{Author Contributions}
RAT, SBS, CJL, YF , WY, and DL implemented the GBT observation campaign. SD and DL implemented the Parkes (Murriyang) observation campaign. RAT, KA, RSL, SBS searched the GBT data for bursts; RAT, LC and KA carried out analysis of their properties. SD searched the Parkes data for bursts and analysed the burst properties. RAT, LC, YF and YKZ conducted the polarization analysis and visualization. RAT, SBS, LC, DL, YF, and SD led the interpretation of the results and writing of the manuscript.  PB, SBS, LC, and RAT produced the turblence model. YPY and BZ contributed theoretical investigation of the physical implications. All authors contributed to the analysis or interpretation of the data and to the final version of the manuscript. 
\section*{Competing interests}
The authors declare no competing interests.
\section*{Data and materials availability}
The observations from the Parkes radio telescope are available from \url{https://data.csiro.au/} after an 18-month embargo period. The Parkes bursts data are available in Science Data Bank at \url{https://doi.org/10.11922/sciencedb.o00069.00007} \cite{parkesdata}.
The GBT bursts data are available in Zenodo at
\url{https://doi.org/10.5281/zenodo.7339930}\cite{gbtdata}.

\section*{Code availability}
Our software and notebooks for the polarization analysis is available at \url{https://github.com/ReshmaAnnaThomas/FRB20190520B} \cite{reshma_anna_thomas_github} and \url{https://github.com/SukiYume/RMS} \cite{sukiyume}. Our derived properties of the bursts are listed in Table S2 for the polarized bursts, and in Table S3 for bursts with no detected polarization.


\section*{Supplementary materials}
Materials and Methods\\
Supplementary Text\\
Figs. S1 to S3\\
Tables S1 to S3\\
References \textit{(42-82)}
\clearpage

\renewcommand{\thesection}{S\arabic{section}}
\renewcommand{\thetable}{S\arabic{table}}
\renewcommand\thefigure{S\arabic{figure}}
\renewcommand{\theequation}{S\arabic{equation}}
\pagenumbering{arabic} 
\setcounter{figure}{0}
\begin{center}
\Huge Supplementary materials for:\\
\Large Magnetic field reversal in the turbulent environment around a repeating fast radio burst  

\vspace{6mm}
\author
{\normalsize
Reshma Anna-Thomas,
Liam Connor, Shi Dai,
Yi Feng, Sarah Burke-Spolaor,
Paz Beniamini,
Yuan-Pei Yang,Yongkun Zhang,
Kshitij Aggarwal, 
Casey J. Law,
Di Li, Chenhui Niu,
Shami Chatterjee, Marilyn Cruces, Ran Duan,
Miroslav D. Filipovi, George Hobbs, Ryan S. Lynch, Chenchen Miao, Jiarui Niu,
Stella K. Ocker, Chao-Wei Tsai, Pei Wang,
Mengyao Xue,  Jumei Yao, Wenfei Yu, Bing Zhang, Lei Zhang, Shiqiang Zhu, Weiwei Zhu}
\\
\vspace{3mm}
\normalsize Corresponding author: E-mail: rat0022@mix.wvu.edu, dili@nao.cas.cn
\end{center}

\baselineskip24pt


\vspace{2cm}
\Large This PDF includes:\\
\normalsize Materials and Methods\\
\normalsize Supplementary Text\\
\normalsize Figs. S1 to S3\\
\normalsize Tables S1 to S3\\
\normalsize References \textit{(42-82)}

\clearpage

\textwidth 16cm 
\textheight 21cm
\footskip 1.0cm%
\baselineskip 24pt

\section{Materials and Methods}
\subsection{Observations of \frb} \label{sec:obs}
A brief summary of all the observations is given in Table \ref{obstable}.
\subsubsection{GBT L-Band}
We conducted L-Band observations of \frb\ ({Right Ascension} 16h02m04.266s, {Declination} -11$^{\circ}$17$'$17.33$''${(J200)}) 2020 September 17 using the 100-m Robert C. Byrd Green Bank Telescope. The observation was done for a total of 6.5 hours on the source. The L-Band {receiver}, Rcvr1\_2, which has a bandwidth of 800 MHz between the range 1100-1900 MHz, was used. The {Versatile GBT Astronomical Spectrometer} (VEGAS) Pulsar Mode backend recorded 8-bit data across the 800 MHz bandwidth in 4096 channels with a frequency resolution of 195 kHz. The native time resolution of the recorded data was 81.92 $\mu$s, and the polarization information of the data was in coherence format, i.e., AABBCRCI ( {where} AA and BB are the direct products of the two input A and B and CR and CI are the real and imaginary parts of the cross product of A and B). The bright quasar J1445+0958 {(14h45m16.46s, 09d58$'$36.07$''$)} was observed for flux calibration in both ON and OFF positions.


\subsubsection{GBT C-Band}
Three sets of GBT C-Band observations were done which we refer to as Epoch 1, Epoch 2, and Epoch 3.
\paragraph{Epoch-1}\mbox{} \\
GBT observations using the C-Band {receiver}, Rcvr4\_6 (4-8 GHz), were conducted on 2020 September 14 for a total of 4 hours. The bright quasar J1445+0958  was observed for flux calibration in both ON and OFF positions, and a noise diode scan for one minute was used in the calibration of Stokes parameters. The 8-bit Full Stokes IQUV data was recorded using the VEGAS Pulsar Mode backend across 12288 frequency channels and was sampled with a time resolution of 87.38 $\mu$s and a frequency resolution of 366 kHz. 

\paragraph{Epoch-2}\mbox{} \\
The second set of GBT observations using the C-Band {receiver}, Rcvr4\_6 (4-8 GHz), was done six months after the first epoch on 2021 March 19,23,27,31. 
The total observing time was 10 hours, with 2.5 hours per day. We did a noise diode scan for one minute to calibrate the Stokes parameters and to verify the calibration procedure, we observed the pulsar PSR B1933+16 {(19h35m47.70s, +16d16$'$40.03$''$)} prior to observing the source. The 8-bit Full Stokes IQUV data was recorded using the VEGAS Pulsar backend across 12288 frequency channels and was sampled with a time resolution of 43.69 $\mu$s and a frequency resolution of 366 kHz. An independent analysis of Epoch 2 is also reported by \cite{yifeng}.

\paragraph{Epoch-3}\mbox{} \\
We did the third round of observations using the C-Band {receiver}, Rcvr4\_6 (3.9-7.3 GHz), on 2021 November 4. The bright quasar 3C 286 was observed for flux calibration in both ON and OFF positions. A 5-minute scan on the pulsar B1933+16 was done prior to the observation of the FRB source. A one-minute noise diode scan to do polarization calibration was done on the test pulsar and the FRB source. The 8-bit Full Stokes IQUV data was recorded using the VEGAS Pulsar Mode backend across 9216 frequency channels and was sampled with a time resolution of 87.38 $\mu$s and a frequency resolution of 366 kHz. {The receptor basis for both GBT L- and C-Band receivers are linear.}


\subsubsection{Parkes ultra-wideband observations}
 FRB~20190520B has been monitored fortnightly at Parkes using the Ultra-Wideband Low (UWL) receiver since 2021 April 5. The UWL system provides radio frequency coverage from $704$\,MHz to 4032\,MHz \cite{2020PASA...37...12H}. Data were recorded with 2-bit sampling every 32\,$\mu$s in each of the 1\,MHz wide frequency channels (3328 channels in total). The integration time of each observation is $\sim7200$\,s. Data were coherently de-dispersed at a DM of 1220.0\,${\rm pc\,cm}^{-3}$ with full Stokes information being recorded. The receptor basis of the Parkes UWL receiver is also linear.

A critical sampling filter bank has been used to produce 26 sub-bands, and we removed 5\,MHz of the bandpass at each edge of the 26 sub-bands to mitigate aliasing. To measure the differential gains between the signal paths of the two voltage probes, we observed a pulsed noise signal injected into the signal path prior to the first-stage low-noise amplifiers before each observation. The noise signal also provides a reference brightness for each observation. To correct for the absolute gain of the system, we use observations of the radio galaxy 3C~218 (Hydra A), using on- and off-source pointings to measure the apparent brightness of the noise diode as a function of radio frequency. Polarimetric responses of the UWL are derived from observations of PSR~J0437$-$4715 {(04h37m15.81s, $-$47d15$'$08.62$''$)} \cite{1993Natur.361..613J} covering a wide range of parallactic angles \cite{2004ApJS..152..129V}, taken during the commissioning of UWL in November 2018. The Stokes parameters are in accordance with the astronomical conventions~\cite{2010PASA...27..104V}. The linear polarization and the position angle (PA) of linear polarization were calculated following \cite{2015MNRAS.449.3223D}. 

\subsection{Data Reduction}
\subsubsection{GBT data}
We used {the software package} \textsc{your} \cite{your} to ingest, pre-process the data and search for single pulses. The data were stored in \textsc{psrfits} \cite{Hotan2004_psrfits} format and were converted to a single total intensity \textsc{filterbank} \cite{filterbank} format using \texttt{your\_writer.py}. A composite {radio frequency interference} (RFI) filter, which uses Savgol \cite{agarwal2019} and Spectral Kurtosis \cite{nita2010} filters with a 4$\sigma$ threshold and Savgol filter window of 15 channels to identify and mask frequency channels, was applied during the conversion process. The RFI mitigated \textsc{filterbank} data was searched for single pulses using \texttt{your\_heimdall.py}, which runs {the package} \textsc{heimdall} \cite{barsdell2012heimdall} on the data. The data was searched {for using} a prior DM range of 1000-1400\,\dmunits\ and maximum boxcar width of 50~ms. Astrophysical bursts were identified from the resulting candidates using the machine learning classifier \textsc{fetch} \cite{fetch2020} and through manual inspection. L-Band data yielded 9 bursts above $\rm S/N > 10$, the brightest one with a $S/N$ of 75. Some of these bursts were also detected by GREENBURST \cite{Agarwal+greenburst+2020}, the realtime detection system at GBT. {Because} most repeaters show narrow banded spectra \cite{aggarwal_121102,Pleunis2021Chime}, we {performed} a sub-banded search on the C-Band data. The total bandwidth was divided into non-overlapping sub-bands of bandwidth 750~MHz and 1500~MHz, and then ran the search pipelines as above. This sub-banded search yielded 6, 2, and 8 bursts in Epochs 1, 2, and 3 of the C-Band observations above a threshold of $\rm S/N >10$. {Several of these bursts had been missed} when the search was done on the 4 GHz bandwidth data. 
 


\subsubsection{Parkes UWL data}
The full UWL band was split into multiple sub-bands for the burst search. We used sub-band bandwidths of 256\,MHz, 384\,MHz, and 512\,MHz to optimize our sensitivity to signals with different characteristic bandwidths. The search was performed using the pulsar searching software package {\sc presto}~\cite{presto} on the high-performance computing facility of the Commonwealth Scientific and Industrial Research Organisation (CSIRO). Strong narrowband and short-duration broadband RFI were identified and marked using the {\sc presto} routine \texttt{rfifind}. We used a 2-s integration time for RFI masking and, the default cutoff to reject time-domain and frequency-domain interference was used in our pipeline. We searched a DM range from 1130 to 1280\,cm$^{-3}$\,pc with a DM step of 0.2\,cm$^{-3}$\,pc. Data were de-dispersed at each trial DMs using the {\texttt{prepdata}} routine with RFI removal based on the RFI mask file. Single pulse candidates with S/N larger than seven were identified using the {\texttt{single\_pulse\_search.py}} routine for each de-dispersed time series and boxcar filtering parameters with filter widths ranging from 1 to 300 samples. Burst candidates were manually examined, and narrowband and impulsive RFI were manually removed. 
The burst bandwidth was measured with the frequency spectrum of each burst.

\subsection{Calibration of bursts}

We made pulse archives for all GBT and Parkes UWL bursts using \textsc{dspsr}\cite{dspsr+vanstraten+bailes} and dedispersed them at the detection DM. 
We then {removed} the frequency channels which did not contain the burst using the \texttt{paz} routine in \textsc{psrchive} \cite{psrchive+hotan}. The calibration data and solutions with respect to frequency channels were visualized using \texttt{pacv} and the channels which gave anomalous solution were {removed}. This typically included the channels at the edge of the frequency band. Then, we self calibrated the data and plotted the Stokes parameters with respect to frequency to check the calibration solutions. The burst archives were calibrated for flux and polarization using the \textsc{psrchive} routine \texttt{pac} in `SingleAxis' mode. In this method, an injected noise diode signal is used to calibrate for the gain and phase of the two polarization channels. This mode, however, assumes that the receptors are orthogonally polarized and the noise signal is equal and at the same phase in both receptors. This method doesn't correct for cross-coupling or leakage. The consistency check for this was done using the PSR B1933+16 observation. The calibrated frequency-averaged profiles of the pulsar were compared to the published profiles in the European Pulsar Network (EPN) archive \cite{hoe1999_B1933+16_EPN}. The bursts in the GBT C-band Epoch 2 were calibrated only for polarization because a flux calibrator was unavailable for that session. 
The polarization of Parkes UWL bursts were calibrated using \textsc{pac} with the Measurement Equation Modeling (MEM) technique\cite{2004ApJS..152..129V}. The polarimetric responses of the UWL are derived from observations of PSR~J0437$-$4715 covering a wide range of parallactic angles and also on-source and off-source observations of
Hydra A (3C~218). These calibration files were provided by the Australian Telescope National Facility.
The on-axis polarimetric response of UWL is described elsewhere~\cite{2020PASA...37...12H}.

\subsection{Quantifying burst properties}

\subsubsection{Rotation Measure}
Rotation measure is defined as the integrated column density of electrons weighted by the magnetic field component along the LOS: 
\begin{equation}
    RM = \frac{e^3}{2\pi m_e^2 c^4}\int_0^d n_e(l)B_{||}(l)dl
\end{equation}


Large RMs can lead to bandwidth depolarization. For finite frequency channel bandwidths $\delta\nu$ at a central observing frequency $\nu_c$, the intra-channel rotation, {$\Delta\theta$,} is given by
\begin{equation}
    \Delta\theta = \frac{RM~c^{2}\delta\nu}{\nu_{c}^{3}} ~\rm rad
\end{equation}
The fractional depolarization,$f_{\rm dpol}$, is then
\begin{equation}
    f_{\rm dpol}=1-\frac{{\rm sin}(2\Delta\theta)}{2\Delta\theta}
\end{equation}
Most repeating FRBs with reported polarimetry are 100\% linearly polarized\cite{MICHILLI2018,Nimmo180916,Kumar2021}. Therefore, {we} assume that \frb\ might intrinsically be 100\% linearly polarized (if observed without LOS propagation effects), and thus for our observed $\sim$50\% polarization, this implies $\Delta\theta=1$\,rad. For our channel bandwidth $\delta\nu$= 0.366 MHz and a center frequency of $\nu_{c}$= 6 GHz, we get a lower limit of the RM that can depolarize the signal in GBT C-Band to be $6.5\times10^{6}$\rmunits.  For our GBT L-Band data with center frequency 1400 MHz and $\delta\nu$= 0.195 MHz, the bandwidth depolarization limit corresponds to $1.5\times 10^{5}$\rmunits.

To calculate the polarization fraction, we de-rotated the burst profiles at their respective RMs, 
and polarized pulse profiles were made by averaging over the frequency dimension. 
In the presence of noise, $L_{\rm meas}=\sqrt{Q^{2}+U^{2}}$ (where $L_{\rm meas}$ is the measured linear polarization from the Stokes parameters $Q$ and $U$) overestimates the linear polarization. Therefore, we calculated the unbiased linear polarization, $L_{\rm unbias}$ \cite{Everett2001}
\begin{equation}
L_{\rm unbias} =
\left\{ 
  \begin{array}{ c l }
    \sigma_{I}\sqrt{\frac{L_{\rm meas}}{\sigma_{I}}-1} & \quad \textrm{if } \frac{L_{\rm meas}}{\sigma_{I}} \geq 1.57 \\
    0                 & \quad \textrm{otherwise}
  \end{array}
\right.
\end{equation}
where $\sigma_{I}$  is the off-pulse standard deviation in Stokes I.

We determine RM using two independent methods, RM synthesis \cite{BrentjensBruyn2005,GHeald2009} and Stokes QU fitting \cite{stokesqufitting}, which yield consistent RM values shown in Table: \ref{mergedtable}.


\paragraph{RM Synthesis}\mbox{}

 We performed RM Synthesis\cite{BrentjensBruyn2005,GHeald2009} on our bursts. RM Synthesis is a non-parametric approach to determining the RM of sources {which uses} a Fourier-like transformation 
\begin{equation}
    F(\phi) = \int_{-\infty}^{\infty}P(\lambda^{2})\exp^{-2i\phi\lambda^{2}}d\lambda^{2}
\end{equation}
where $\lambda$ is the observing wavelength, $F(\phi)$ is  the Faraday dispersion function (FDF), $\phi$ is the Faraday depth and $P(\lambda^{2})$ is the total linearly polarized flux intensity. This method assumes that the source is a sum of emitters at different Faraday depths. The millisecond duration of FRBs implies that the emission arises from a compact region. This restricts the amount of differential Faraday rotation that can happen in the emission region, {which} can be considered as a Faraday thin source. In the Faraday thin regime, the FDF peaks at a single value of Faraday depth ($\phi$) which is essentially similar to the RM. 
{We performed 1-D RM synthesis on the burst data}  using the package \textsc{rm-toolkit}.

\paragraph{Stokes QU Fitting}\mbox{}

Another way of determining RM is to fit {a sinusoidal model to} the oscillations in Stokes Q and U as a function of $\rm \lambda^{2}$.
The {model} fitting includes two parameters: the RM and the polarization angle at infinite frequency, $PA_{\infty}$. Optimal parameters were determined numerically through Markov Chain Monte Carlo methods \cite{sukiyume}. Parameter estimation seeks to optimize the likelihood function given a model and the data. The prior logarithmic likelihood $\Lambda$ is:
\begin{equation}
    \log\Lambda = -\frac{1}{2}\sum_{i=1}^{N}\{\frac{(Q_i-L_i \cos\theta_i)^2}{\sigma_{Q_i}^2}+\frac{(U_i-L_i \sin\theta_i)^2}{\sigma_{U_i}^2}\},
\end{equation}
where $N$ is the number of frequency channels, $\sigma_{Q,U}$ is the single channel RMS, and $\theta_i = 2 (PA_{\infty}+RM \lambda_{i}^2)$. $L_i$ is given by:
\begin{equation}
    L_{i} = \frac{\frac{Q_i\cos\theta_i}{\sigma_{Q_i}^2}+\frac{U_i\sin\theta_i}{\sigma_{U_i}^2}}{\frac{\cos^2\theta_i}{\sigma_{Q_i}^2}+\frac{\sin^2\theta_i}{\sigma_{U_i}^2}}.
\end{equation}




The RM values detected for the bursts by both the methods and the polarization fraction obtained by de-rotating the bursts at respective RM are listed in Table~\ref{mergedtable}, {along with the}  $1 \sigma$ {uncertainity in} the polarization fraction. For the bursts without an RM detection,  Table\,\ref{DM_pol_table} lists the 3$\sigma$ upper limit of polarization fraction at 0 \rmunits. The values of the RM were then converted from the observer's frame, $RM_{\rm obs}$ to the source frame,$RM_{\rm src}$, by 
\begin{equation}
    RM_{\rm src} = RM_{\rm obs}\times(1+z)^{2},
\end{equation}
where $RM_{\rm src}$ is the RM in the source frame, and $RM_{\rm obs}$ is the RM in the observer's frame of reference.

\subsubsection{Verifying calibration with pulsar data}

In Epochs 2 and 3 of GBT observations, PSR B1933+16 was observed adjacent to our observations to serve as an additional calibration test. Two instances of the pulsar were observed in Epoch 2 (one adjacent to burst C2.1 and the other adjacent to burst C2.2). We verified our calibration and RM search methods by calibrating the pulsar B1933+16 for Epoch 2 and Epoch 3 using the same procedures applied to the burst data.  We measured the RM of the average pulse profile (folding the data at the period of the pulsar using the software \textsc{dspsr}\cite{dspsr+vanstraten+bailes}) and carried out an RM search using RM synthesis.

In all of the pulsar observations, the polarization and RM properties were internally consistent and agreed {with previously reported values \cite{hoe1999_B1933+16_EPN}. }The previously published RM of this pulsar is ${\rm RM}= -10.2\pm3\,{\rm rad\,\mathrm{m}^{-2}}$ \cite{hoe1999_B1933+16_EPN}. The RMs we measured in the three observations, chronologically, were $-54\pm 117\,$\rmunits, $-64\pm 121$\,\rmunits\  and \mbox{$-77\pm 91$\,\rmunits,} respectively. The consistency of these values with one another and with the past published RM value {gives us} confidence in the RM swings we report for \frb. {The} sign {appear consistently in the independent} GBT and Parkes observations of \frb. {We also consider it unlikely that there was large} sporadic instrumental-based RM variability at both GBT and Parkes. 


\subsubsection{Dispersion measure}
The mean pulse DM {previously} reported for \frb\ is 1204$\pm$4 \dmunits\cite{Niu2021}. 
 FRBs have intrinsic narrow-band frequency structures that are preferentially ordered in time from high to low frequency \cite{Hessels2019}. Such structures are visible by eye for some bursts, as in Figs.~\ref{sgram} and \ref{fig:dedisp}. Determining DM by maximizing the frequency-integrated S/N can over- or under-estimate the DM when these structures are present. Therefore, we report only the structure- maximizing DMs for our bursts.



\paragraph{Structure Maximizing DM}\mbox{}

For all bursts in the GBT and Parkes sample, we report the {DM value that} maximizes the frequency averaged burst structure. This method requires high S/N and/or sharp temporal components in the pulse{; it is less accurate} for low S/N bursts or bursts without time-resolved structure. We use the package \texttt{DM\_PHASE} \cite{Seymour2019}, which computes the structure-optimized DM of a burst by maximising the coherent power across the bandwidth. The uncertainities are {calculated as} the standard deviations in DM {produced} by converting the standard deviations in the coherent power spectrum using a Taylor series \cite{Seymour2019}. Measured this way, the burst DMs fall in a range between 1170 to 1259\,\dmunits, with a 
mean of 1206.6\,\dmunits\ and a standard deviation of 13.6 \dmunits. These values are consistent with observations reported at other epochs {and in previous} works\cite{Niu2021,yifeng}.

\paragraph{DM Variability}\mbox{}

Given the small uncertainities in DM and the large variance in the structure-maximizing DM measurements, there is evidence of DM changes between bursts. We inspected the DM {of} bright bursts {fitted} automatically and by hand. {Figure.~\ref{fig:dedisp} shows the brightest and most highly structured bursts as examples, which have been} dedispersed at low, average, and high-range DM values {output} by the software. We {identify} DM variability {of about} 10\,\dmunits\ {in the} bursts in {the first, second and fourth row}, which each have complex substructures aligned at DM values visually separated by approximately 13\,\dmunits. {This is} similar to the spread that appears in the structure-maximized DM measurements as reported in Table \ref{DM_pol_table} and plotted in Fig.~1. Thus, while some of the scatter could be due to intrinsic (non-DM) sweeps, there is evidence of {variability} in DM.


The average LOS magnetic field integrated through the Faraday medium {can be} estimated from the RM and DM variance as
\begin{equation}
    <B_{||}> =1.23 ~\frac{\Delta RM}{\Delta DM}~\rm \mu G.
\end{equation}
Using 
$\Delta DM\simeq 10\,$\dmunits and a net 
$\Delta RM_{\rm src}\simeq 57\times10^{3}$, we get
$<B_{||}> \simeq 5.7 {\rm mG}$.

This {calculation assumes} that the RM and the DM changes are happening in the same LOS medium. {Because} the medium is turbulent, we assume that $\rm \Delta DM \sim <DM>$ within the screen and the non-fluctuating component of the DM in the screen is negligible. 

The fractional variation of the LOS magnetic field component is given by:
\begin{equation}
    \Delta B_{||} = 1.23 \left ( \frac{\Delta RM}{DM} + \frac{RM\Delta DM}{DM^{2}} \right ) \rm\mu G.
\end{equation}
Because the second term is small compared to the first, we can write 
\begin{equation}
    \frac{\Delta B_{||}}{B_{||}} \approx \frac{\Delta RM}{RM}
\end{equation}


\subsubsection{Detailed Measurements}
The properties of all bursts with detected polarization are shown in Table S1. 

We used \textsc{burstfit} \cite{aggarwal_121102} to model the burst profiles of all the bursts, 
using a Gaussian function and the non-linear least-squares fitting implemented in \texttt{scipy.curve\_fit}. {There were three free parameters:} fluence, width, and the location of the profile. The {best fitting values and their uncertainties} are shown in Table \ref{mergedtable}.


\section{Supplementary text}
\subsection{Scattering and scintillation}
The pulse intensity is fully modulated {in frequency}, indicating that it is spatially resolved by a scattering screen in the Milky Way. We fitted a Lorentzian profile to the autocorrelation (ACF) function {of the time-averaged frequency spectrum} of our brightest bursts, finding a decorrelation bandwidth of $\Delta\nu\approx35\pm$1\,MHz. This is consistent with our 1.4\,GHz measurement of scintillation ($\Delta\nu\approx$\,0.5\,MHz) and previous measurements \cite{Niu2021,ocker+22}, as well as the value predicted by the NE2001 free electron model \cite{ne2001}. The presence of Galactic scintillation 
and temporal scattering suggests the latter is local to the source, because 
the Milky Way's screen would resolve the scattered pulse if the screen were in an intervening galaxy since angular broadening would be large \cite{masui2015}.  

\subsection{Comparison with other astrophysical sources}

RM variations have been reported in other FRBs and pulsars. FRB\,20121102A (05h31m59s, +33d08$'$53$''$) has a decreasing trend in RM, varying about 200\,rad\,m$^{-2}$\,day$^{-1}$. It also shows short time scale variations of $\sim 10^{3}$\,rad\,m$^{-2}$\,week$^{-1}$ \cite{Hilmarsson121102}. This decrease was interpreted as an expanding nebula near a supernova remnant or due to the source being in the vicinity of a massive {black hole} \cite{Hilmarsson121102}. However, the young supernova model {interpretation has been disputed\cite{Plavin2022}}.
\frb\, is distinct from FRB\,20121102A in its large host DM, local scattering, large fractional RM variation, and its magnetic field sign reversal.

FRB\,20201124A (05h08m3s, +26d03$'$38$''$) shows irregular short-term variation in RM, followed by a period of steady RM\cite{Xu2022}. This source also shows RM variation within the duration of a burst. {A magnetar/Be star binary model was proposed to explain this RM variation \cite{Wang2022,Zhao2023}}. RM variations have also been seen in FRB\,20180916B (01h58m01s, +65d43$'$00$''$), where a period of small-scale fluctuation in RM (\rm$\Delta \rm RM\sim2-3$~\rmunits \cite{Pleunis2021lofar}) is followed by a period of secular increase \cite{McKinven2022_180916RM}. The magnetar J1745--2900 (17h45m40s, --29d00$'$30$''$), {located in the Galactic Center,} shows a variation of about $\rm 7.4$\,rad\,m$^{-2}$\,day$^{-1}$ and small DM changes \cite{eatough-2013, desvignes+18}. This variation was attributed to variation in the projected magnetic field along the line of sight of the rapidly rotating magnetar {and} depolarizes rapidly towards low radio frequencies. RM variations {have been} reported in the pulsar binary system PSR B1744$-$24A (17h48m02s, --24d46$'$37$''$), {located in a globular cluster, which exhibits} irregular RM variations at random orbital phases as the pulsar passes close to the companion \cite{Dongzi2022}. 

The pulsar PSR B1259$-$63 (13h02m48, --63d50$'$09$''$) is in a binary system with a Be star causing RM variations during its periastron passage. Away from the periastron, the RM of the pulsar is $\sim 21$~\rmunits. But during the periastron passage, the RM shows several orders of magnitude variations and sign reversals\cite{Johnston1996,Connors2000,Johnston2005}.

\subsection{Interpretation}
{We interpret} the spectral dependence of the polarization, the strongly fluctuating values of RM, and the strong scintillation (characterized by a {ms-level} scattering time) measured in the 1\,GHz band {as due to} multi-path propagation. 
{This requires} two basic components: First, a highly turbulent plasma, whose turbulence properties sets the variance timescales in RM, DM, and scattering. Second, a large magnetic field, e.g. {the} bulk magnetic field externally generated from a star or large variation in a magnetic field generated in the turbulent plasma medium via the action of ordered shocks from some external source. The magnetic field sets the observable range of RM. In the binary scenario, the {range of} RM variance {is expected to} change periodically (on the orbital timescale), while in a non-binary scenario (e.g., a view of an FRB through turbulent shocks), the fluctuations would continue on the same scale {indefinitely}.

We {develop a model of} a screen with parameters that satisfy the various observational constraints.  A turbulent medium has eddying structures on various spatial scales; we consider plasma as the turbulent medium. 
Strong scintillation implies that the diffraction scale of the plasma screen (over which the phase of an incoming wave changes by {360$^\circ$}) is very small relative to the Fresnel radius. The size of patches over which the PA changes by {a full cycle}, $\ell_{\chi}(\nu)$, is expected to be much larger \cite{2021arXiv211000028B}. If $\ell_{\chi}(\nu)$ is approximately the refractive scale (i.e., the observable size of the screen) at {a frequency given by} $\nu_{\chi \rm s}$, then above this frequency, the polarization will remain large (and there will be little circular polarization), while at lower frequencies the polarization will be strongly reduced (and there will be circular polarization of tens of percent of the linear polarization). The observed depolarization at the L-band, combined with the {polarization rising almost to 100\% polarization for some bursts measured at  C-band,} and the tight upper limits on the circular polarization component at the latter band, imply $\nu_{\chi \rm, s}\sim 5$GHz. The variability of RM by \mbox{$\delta\mbox{RM}\sim 300\,$\rmunits} over a scale of less than one week dictates the value of $\ell_{\chi}$. For convenience, we use the scaling variable  \mbox{$\delta\mbox{RM}_{2.5}\equiv \delta\mbox{RM}/10^{2.5}$\rmunits} to represent the RM variability on timescales of one week.

The first constraint on the screen properties comes from the measured fluctuations in RM and DM between different bursts. Attributing these differences to the properties of the fluctuating screen, we estimate 
\begin{equation}
    B_{||}\approx 1.2 \Delta \mbox{RM}_{4} \Delta \mbox{DM}_{1}^{-1}~{\rm mG}
\end{equation}
and 
\begin{equation}
    L\approx 10 n_e^{-1}\Delta \mbox{DM}_{1}\mbox{pc}
\end{equation}
where $B_{||}$ is the component of the magnetic field within the screen parallel to the line of sight, $L$ is the thickness of the screen, $n_e$ is the {electron} density (number per cm$^{-3}$) and $\Delta \mbox{RM}_4\equiv \Delta \mbox{RM}/10^4\mbox{rad m}^{-2}$, $\Delta \mbox{DM}_1\equiv \Delta \mbox{DM}/10\mbox{pc cm}^{-3}$ are the mean dispersion measure and rotation measure accumulated through the screen. The turbulent nature of the screen implies that $\Delta \mbox{RM}$ ($\Delta \mbox{DM}$) also sets the degree of variability in RM (DM) that can be seen over sufficiently long timescales (over long enough times, the eddies on all scales will have moved/rotated, {so} their summed contributions randomize). All other properties of the turbulent screen can be {described by three} more physical quantities: $d,l_{\rm max}$ (respectively describing the distance of the source to the screen and the largest turbulent eddy size) and $v_{\rm max}$ (the maximum {velocity} between the FRB {source's} proper motion, the eddies' turnover velocity, and the velocity of the screen relative to the line of sight) the value of which is {well-}constrained a priori, and is expected to be $\sim 10^2-10^3 \mbox{km s}^{-1}$ {considering a typical natal kick for a neutron star, e.g. \cite{lynelorimer}.} There are three additional observational constraints: (i) the scattering time measured at L-band (This is strictly an upper limit on the scattering time associated with the magnetized screen accounting for depolarization, because it is possible that the screen dominating the scintillation is physically distinct from {the scattering screen}.), (ii) $\nu_{\chi \rm, s}\sim 5$GHz (see above) and (iii) $\ell_{\rm \delta RM,2.5}/v_{\rm max}< 1$\,week (where ${\rm \delta RM},2.5\equiv {\rm RM}/10^{2.5} \mbox{rad m}^{-2}$, $\ell_{\rm \delta RM,2.5}$ is the scale over which the screen RM changes by $10^{2.5}\mbox{ rad m}^{-2}$ and $\ell_{\rm \delta RM}=(\delta \mbox{RM}\lambda^{2})^{6/5}\ell_{\chi}$).  These conditions can be re-written as
\begin{eqnarray} \label{eq:paz1}
&\frac{L}{\ell_{\rm max}}< 1.6\times 10^7\nu_{\rm co,2}^{-5/4} \Delta \mbox{DM}_{1}^{-7/4} n_1^{-5/4}\\
&\frac{L}{\ell_{\rm max}}=5\times 10^{8} \left(\frac{d}{L}\right)^{-5/4} \nu_{\chi, \rm s,5GHz}^{23/4} \Delta \mbox{DM}_{1}^{-1/4}\Delta \mbox{RM}_4^{-3/2}n_1^{-5/4}\\
&\frac{L}{\ell_{\rm max}}> 1.5\times 10^7\Delta \mbox{DM}_{1}^{5/2}\Delta \mbox{RM}_4^{-3} n_1^{-5/2}v_{\rm max,8}^{-5/2}t_{\rm w}^{-5/2} \delta RM_{2.5}^{-3}\label{eq:paz3}
\end{eqnarray}
where {$\nu_{\rm co,2}\equiv 100$Hz} is the decorrelation bandwidth inferred at L-band), $\nu_{\chi, \rm s,5\mbox{GHz}}\equiv \nu_{\chi,\rm s}/5$GHz, $n_1 \equiv n_e/10\mbox{cm}^{-3}$, $v_{\rm max,8}\equiv v_{\rm max}/10^8 \mbox{cm s}^{-1}$, $t_{\rm w}\equiv t/1\mbox{ week}$ and $\delta {\rm RM_{2.5}}$ is the fluctuation in the RM on the timescale $t_{\rm w}$, as previously defined.
A range of solutions for these equations exist for our set of observations if $2\mbox{ cm}^{-3}< n_e< 5\times 10^6\mbox{ cm}^{-3}$, $10^{-5} < d < 100\mbox{ pc}$, $20< d/L < 3\times 10^4$. {(Note that our lower limit on $d$ is equivalent to about 8~astronomical unit (au)).If the LOS $\Delta {\rm DM}$ is actually an order of magnitude smaller, the allowed values become $5\times10^{-3}\,\mbox{cm}^{-3}< n_e< 1\times 10^{10}\mbox{ cm}^{-3}$, ${3\times10^{-9}\,\mbox{pc}}< d < 300\mbox{ pc}$, and $1.5< d/L < 10^6$.}
For such a screen, we predict the spectral dependence of the linear polarization fraction would scale as $\nu^{1.7}$ below $\approx 1.5\mbox{ GHz}$ (with a slightly steeper dependence between this frequency and $\nu_{\chi \rm s}$,  consistent with the observed polarization change between the L- and C-band). Alternatively, at frequencies higher than the C-band, we expect the depolarization to be negligible. 
The timescale associated with stochastic changes in the polarization angle and the intensity due to the scintillating screen is related to the motion relative to the line of sight of the diffraction scale \cite{2020MNRAS.498..651B}. The latter can be many orders of magnitude smaller than $\ell_{\chi}$. For the same parameters as above, this timescale is $10^{-4}<t_{\rm var}<10^{-1}\mbox{ s}$.

What could be the origin of such a screen? Generally, any extended plasma region with a magnetic field {at least several mG} that is highly turbulent. The most common  physical scenario that would allow this configuration would be an active wind seen at a line of sight that passes close to a foreground object (e.g., an FRB source seen through a stellar wind, as discussed in the main text). However, we outline other scenarios below.

{\bf Shocked neutron-star (NS) wind}. One possibility is that the central object is a magnetized NS and that the magnetized screen is provided by the shocked NS wind that lies downstream of the termination shock. The central NS can supply the magnetic field in this scenario. The required $B\approx 1$\,mG implies a screen distance of $d_{\rm B}=B_{\rm NS} R_{\rm NS}^3 \Omega^2/(B c^2)$, where $B_{\rm NS}$ and $R_{\rm NS}$ are the magnetic field and radius of the NS, and $\Omega$ is the NS spin frequency. 

We assume that the field decays as a dipole at large distances up until the light cylinder, beyond which $B\propto r^{-1}$. A necessary condition is that the required radius {falls} above the termination shock, $r_{\rm s}\approx \sqrt{B_{\rm NS}^2 R_{\rm NS}^6 \Omega^4/(4\pi P c^4)}$ (where $P$ is the ram pressure). The requirement $d_{\rm B}>r_{\rm s}$ is independent of the NS properties and depends only on the ratio of energy densities at the termination shock: $d_{\rm B}/r\approx 4\pi P/B^2$. For $B=1$mG, it translates to $P>8\times 10^{-8}\mbox{ erg cm}^{-3}$, which is plausible in an interstellar medium environment. A major obstacle for this scenario is that the shocked NS wind is typically too tenuous to provide the measured $\Delta \mbox{DM}$. As an example, for $B=10^{14}$\,G, $\Omega=100$\,Hz, $d_{\rm B}\approx 0.4$\,pc. At this {screen distance}, the density of the shocked wind powered by the spindown luminosity is $n_e\sim 4\times 10^{-3}\gamma_{\rm w}^{-2}d_{\rm pc}^{-2}\mbox{ cm}^{-3}$, where $\gamma_{\rm w}$ is the Lorentz factor of the wind (which {could be much greater than one}). The excess DM that this wind can provide is therefore $< 2\times 10^{-4}\mbox{pc cm}^{-3}$, much smaller than the observed value. 

We have assumed an electron-positron plasma which maximizes the density for a given luminosity but also implies no Faraday rotation (and therefore no depolarization by the screen); {Such an electron-positron wind is the standard wind from a pulsar (or magnetar). This is the result of the strong electric fields in which the particles are accelerated, which leads them to develop very high energies and then create high energy photons that produce electron-positron pairs \cite{goldreichjulian69}}. So for such a wind to provide a viable screen, there must be some baryons in the flow, reducing the estimated density. {Alternatively, the shocked density wind could be powered} by episodic mass ejections from the neutron star. Such mass ejections could contribute a much higher density (and the necessary baryons) to the shocked wind. This situation relies on the magnetar having gone through a giant flare at a time $\sim d/c$ prior to the FRB emission, which for the example above is several years.

A second possibility, {which does not rely on recent} mass ejections, {is that the magnetar shocked wind could mix at a} contact discontinuity with the shocked circum-stellar medium (CSM) above it (e.g., due to Rayleigh-Taylor instability). If such mixing occurs efficiently, the shocked CSM could provide the necessary density, while the shocked wind provides the magnetic field. If the NS is sufficiently young, the surrounding material the wind is pushing into could be dominated by the  supernova remnant, which {would provide a high} density.

{\bf Material around a massive black hole} The radio-loud magnetar J1745--2900 {(17h45m40s,-29d00$'$30$''$), located} 0.12\,pc from Sgr A$^*$\cite{eatough-2013}, is a Galactic example of a high-RM source with observed variability \cite{desvignes+18}. Its DM 
is large but stable.
{The magnetar} is 
also depolarized at low frequencies and strongly scattered. If the compact PRS associated with \frb\ were 
due to accretion onto an intermediate-mass black hole (IMBH), 
a dense, turbulent outflow or magnetized accretion disk might account for the 
FRB's time- and frequency-dependent polarization 
effects. The prompt radio emission could be produced 
by a nearby neutron star \cite{pen-connor}, as with J1745--2900, 
or the accretion disk itself \cite{2017ApJ...842...34W,2017MNRAS.471L..92K,Sridhar2021}. 

An IMBH accreting close to {(or above)} the Eddington limit would drive powerful outflows. The density of the outflow $n_w$ at distance $d$ can be estimated as $n_{\rm w}\approx \dot{M}_{\rm w}/(4\pi d^2 v_{\rm w})\approx 2 (M/M_{\odot})f_{\rm M} d_{15}^{-2}v_{\rm w,0.1c}^{-1}\mbox{ cm}^{-3}$, where {$M$ is the IMBH mass and} $\dot{M}_{\rm w}$ is the wind outflow rate, {and as a fraction of the Eddington-limit accretion rate $\dot M_{\rm edd}$ we can define} $f_{\rm M}\equiv \dot{M}_{\rm w}/\dot{M}_{\rm edd}$. {We have normalized here by a wind velocity of $v_{\rm w}\approx 0.1c$.} Considering the constraint $d/L\geq 20$, the observed excess DM of the screen can be reproduced if the screen is placed at a distance $d< 3\times 10^9 (M/M_{\odot})f_{\rm M}\mbox{ cm}$. Imposing the condition $d> 10$au {based on an order-of-magnitude approximation to our lower limit on the screen depth}, we find $M>5\times 10^4 M_{\odot} f_{\rm M}$. In the case of FRB\,20121102A, a multi-wavelength 
analysis of the associated PRS 
concluded that it was more likely to be a neutron 
star wind nebula than an active galactic nucleus (AGN) \cite{chen2022}. 

High RMs have been observed in some {active} galactic nuclei (e.g., toward Sgr A* \cite{marrone07} and other active galactic nuclei \cite{hovatta12}). 
RM reversals have been observed in one source {(1226+023/3C~273),} ranging over a few hundred \rmunits\ \cite{hovatta12}. However, polarization and rotation measurements for AGN cores are often affected by competing contributions of multiple confused sources (or extended structures) that pass through different lines of sight, thus possibly obfuscating RM measurements \cite{hovatta12}. Transients in the same environments have high luminosity from an extremely compact region, therefore providing effectively one sightline.

\subsubsection{Binary model of PSR B1259--63}
For comparison, we also applied the binary model to PSR B1259--63.
For this analysis, the main constraints are: (i) the coherence band at 5 GHz near periastron is around $<$20Hz, (ii) the depolarization frequency is $\sim$5 GHz, (iii) the screen's RM {changes} by 2200\,${\rm rad~m^{-2}}$ within a day, and (iv) the scintillation (diffractive) timescale is $<$15 minutes ({allowing that the plasma} screen dominating scintillation could be distinct from the one causing depolarization) \cite{Johnston1996,Johnston2005}. 
It is possible to satisfy all these criteria simultaneously with a screen ${\rm RM}\simeq3\times10^4\,$\rmunits, screen ${\rm DM}\simeq10\,$\dmunits, and a screen velocity of $\sim$100 km\,sec$^{-1}$. 
Permitted solutions to Equations \ref{eq:paz1}-\ref{eq:paz3} are in the range $10^5\,{\rm cm^{-3}}<n_e<10^7 {\rm cm^{-3}}$, $0.3\,{\rm au }<L<30\,{\rm au}$, which is within the range {expected} for Be star accretion discs discussed by \cite{bestar}.


\clearpage

\begin{figure}[h!] 
 \centering
  \includegraphics[width=0.75\textwidth]{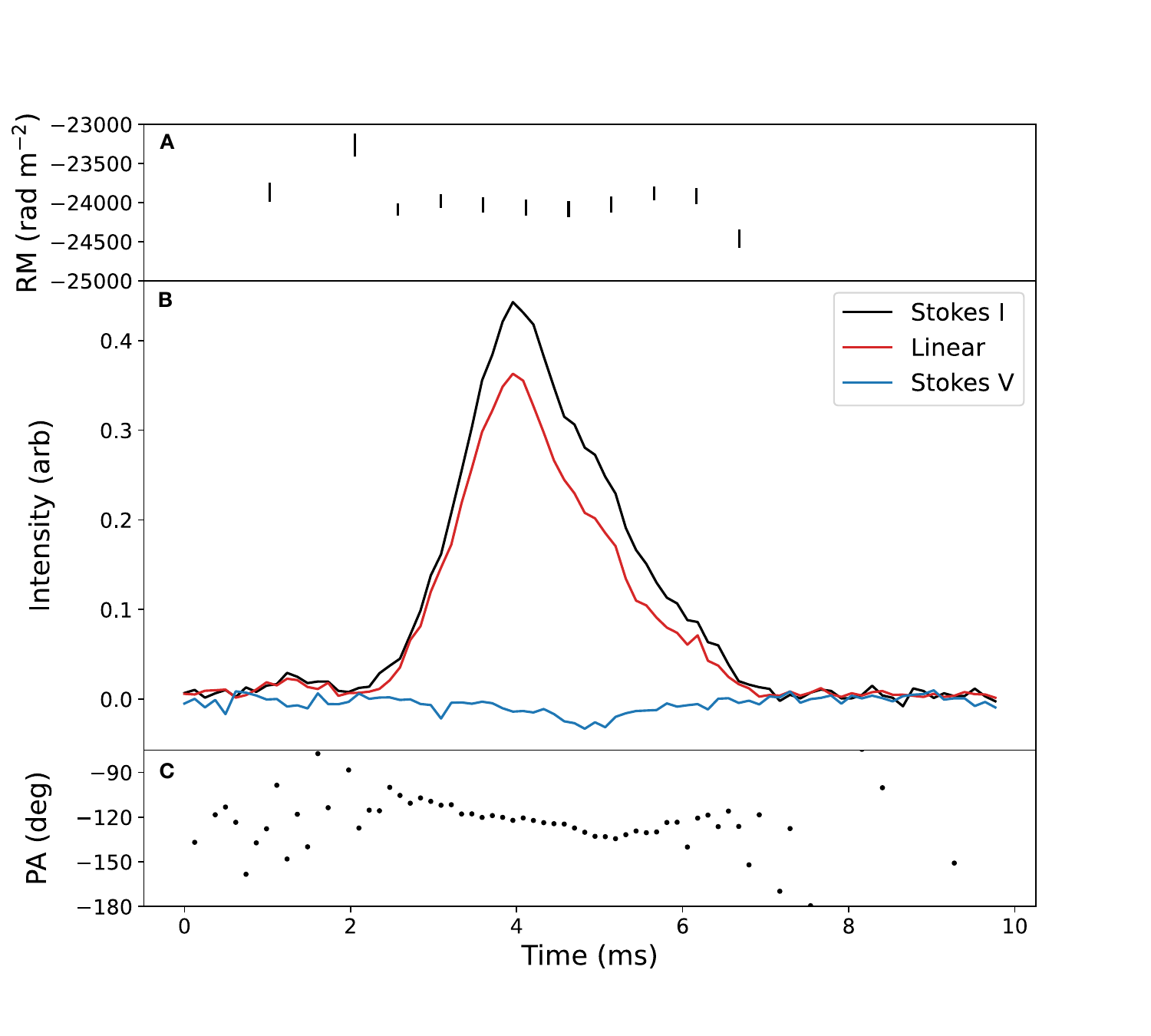}
  \caption{\textbf{The polarized pulse profile of the brightest 4-8\,GHz burst (pulse number C3.2) in our sample}. The top panel (A) shows RM vs. dedispersed pulse arrival time. Pulse profiles in Stokes I, L, and V are shown in the middle panel (B) . The bottom panel (C) is polarization position angle (PA) as a function of time. 
  We find no evidence for RM variation 
  over the pulse, significant circular polarization, 
  nor large PA change across 
  the pulse.}
 \label{fig:bright-cband-burst}
\end{figure}

\begin{figure*}[h!]
    \centering
    \includegraphics[width=0.9\textwidth]{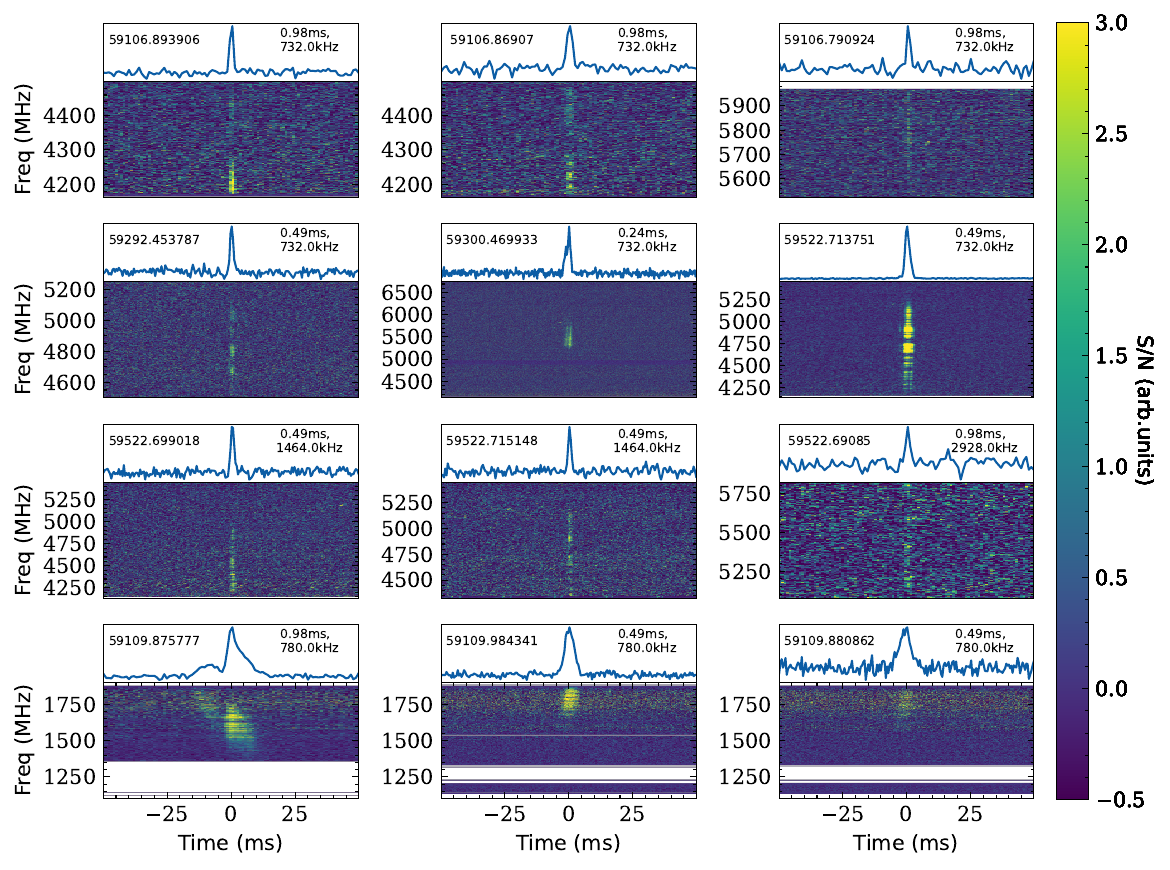}
    \caption{\textbf{The burst spectrograms for a subset of {bright} bursts in our GBT sample.} The dynamic range of the spectrograms was limited to 3 times the standard deviation of the spectrogram for clarity. Data was {downsampled} in time and frequency for better visualization. The time and frequency resolution of data is shown in the top right of the plot. The burst MJDs are given at the top left corner of the plot. }
    \label{sgram}
\end{figure*}

\begin{figure*}[h!]
    \centering
    \includegraphics[width=0.9\textwidth]{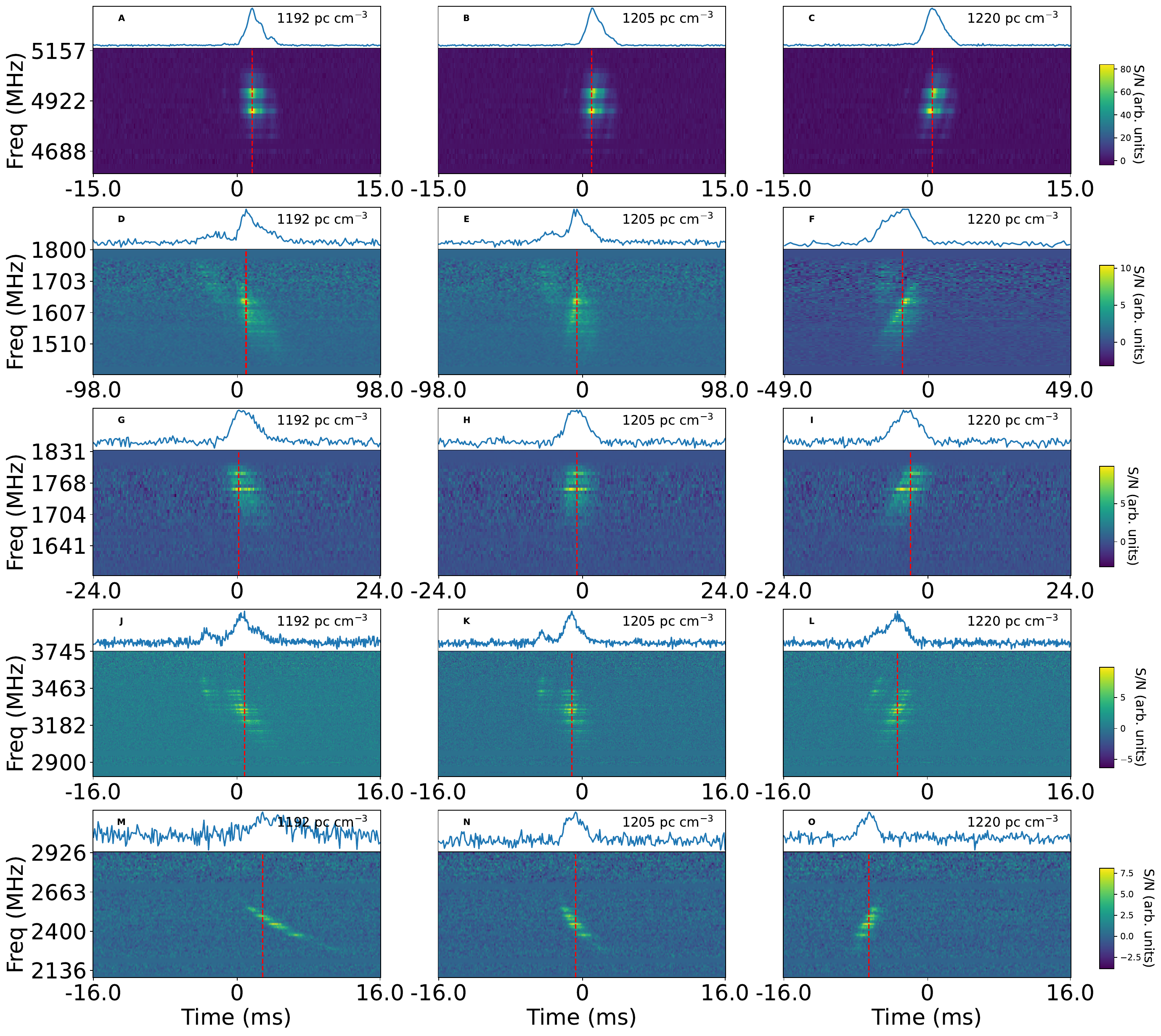} 
    \caption{\textbf{Spectrogram and frequency-averaged profile of a sample of bursts dedispersed at three different DM values}. The first column represents the structure maximizing DM of burst C3.2 (the highest S/N burst in the sample), the second column represents a DM of 1205 \dmunits (average DM of the sample) and the third column represents a DM of 1220 \dmunits. Each row represent different bursts. The red vertical line is a visual reference to compare with different values of the dispersion correction.}
    \label{fig:dedisp}
\end{figure*}

\clearpage

\begin{table}
\begin{center}
\begin{tabular}{|c|c|c|c|c| } 
 \hline
  \textbf{Observation} & \textbf{Start MJD} & \textbf{$\nu_{c}$} & \textbf{$\Delta\nu$} & \textbf{$\Delta t$} \\ 

 & &(MHz) &(MHz)& ($\mu s$) \\
 \hline
 GBT L-Band &59109  & 1400 & 0.195 & 81.92 \\ 
 \hline
  & 59106 & 6000 & 0.366 &87.38\\ 
          & 59292 & 6000 & 0.366 &43.69\\
GBT C-Band       & 59296 & 6000 & 0.366 & 43.69 \\
         & 59300 & 6000 & 0.366 & 43.69 \\
         & 59304 & 6000 & 0.366 & 43.69 \\
  & 59522 & 5637.5 & 0.366 & 87.38 \\
\hline
 & 59332 & 2368 & 1.0 & 32 \\
 & 59351 & 2368 & 1.0 & 32 \\
 & 59360 & 2368 & 1.0 & 32 \\
 & 59373 & 2368 & 1.0 & 32 \\
 & 59384 & 2368 & 1.0 & 32 \\
 & 59400 & 2368 & 1.0 & 32 \\
 & 59415 & 2368 & 1.0 & 32 \\
PKS UWL & 59430 & 2368 & 1.0 & 32 \\
 & 59453 & 2368 & 1.0 & 32 \\
 & 59473 & 2368 & 1.0 & 32 \\
 & 59481 & 2368 & 1.0 & 32 \\
 & 59488 & 2368 & 1.0 & 32 \\
 & 59496 & 2368 & 1.0 & 32 \\
 & 59562 & 2368 & 1.0 & 32 \\
 & 59574 & 2368 & 1.0 & 32 \\
 & 59588 & 2368 & 1.0 & 32 \\
 \hline
\end{tabular}
\end{center}
\caption{\textbf{List of all observations of \frb.} Start MJD is in {the} topocentric reference frame. The center frequency of the observing band, the frequency resolution and time resolution are also given.}
\label{obstable}
\end{table}
\clearpage
\begin{landscape}
\small   
\setlength{\tabcolsep}{2.5pt}
\LTcapwidth=1.2\textwidth
\begin{longtable}{cccccccccccc}

\caption{\textbf{Properties of all bursts of FRB 20190520B with RM detections.} \\
The burst ID's that start with C are detected by GBT C-Band and those that start with P are
detected by Parkes UWL.\\
MJDs are in barycentric dynamical time (TDB) and are referenced to infinite frequency.\\
$S/N$ is the detection signal-to-noise reported by the single pulse search software.\\
$f_{low}$ and $f_{high}$ are the minimum and maximum frequency extent of the bursts, determined visually. \\
$S$ is the fluence of the burst along with $1\sigma$ uncertainity.\\
$W$ is the width in milliseconds along with $1\sigma$ uncertainity. \\
$DM_{stru}$ is the structure-maximizing DM reported by \texttt{DM\_PHASE} with $1\sigma$ uncertainities. \\
$RM_{FDF}$ are the RM values (observer's frame) reported by the RM Synthesis method. \\
$ RM_{QUfit}$ are the RM values (observer's frame) reported by the Stokes QU fitting method. \\
$L/I$ is the percentage of unbiased linear polarization  along with $1\sigma$ uncertainity\\
$V/I$ is the percentage of circular polarization  along with $1\sigma$ uncertainity.}  
\\
\hline
    \  ID& MJD  &  $S/N$  &  $f_{low}$ &   $f_{high}$ &  $S$  & $W$ &  $DM_{stru}$ & $RM_{FDF}$ & $RM_{QUfit}$ & $L/I$  & 
    $V/I$ \\
    & &  &  (MHz) &   (MHz) &  (Jy\,ms) &  (ms) &  (${\rm pc~cm^{-3}}$) & (${\rm rad~m^{-2}}$) & (${\rm rad~m^{-2}}$) & $\%$  & $\%$ \\
\hline \\
\hline

      C1   & 59106.79092398 & 10 & 5500 & 6000 & $0.20\pm0.20$ & $1.10\pm0.10$ & $1224.0\pm0.4 $  & $-7864^{695}_{695}$& $-8222^{360}_{322}$ &   $50.0\pm10.0$ &   $15.0\pm6.0$ \\
      C2.1 & 59292.45378713 & 17 & 4500 & 5250 & $0.34\pm0.02$ & $1.36\pm0.09$ & $1216.4\pm0.2$& $3763^{139}_{139}$    &    $3888^{110}_{104}$ &   $60.0\pm10.0$ &   $-42.0\pm7.0$\\
      C2.2 & 59300.46993316 &               18 &     4117 &       6750 &  $0.36\pm0.01$ &     $2.17\pm0.09$  & $1184.7\pm0.3$ & $1858^{79}_{79}$& $1613^{47}_{54}$ &      $45.0\pm6.0$ &    $13.0\pm2.0$ \\
      C3.1 & 59522.69085019  &               10 &     5075 &       5825 &  $0.17\pm0.02$ &      $1.50\pm0.20$ & $1215.8\pm0.9$ &   $-23257^{364}_{364}$ &       $-22878^{2}_{10}$ &   $60.0\pm20.0$ &    $24.0\pm8.0$ \\
      C3.2 &59522.71375078 &              161 &     4133 &       5450 &  $3.97\pm0.03$ &$2.02\pm0.02$ &  $1191.9\pm0.1$ & $-24163^{52}_{52}$ & $-23997^{8}_{7}$ &   $82.0\pm1.0$ &  $6.0\pm7.0$ \\
      
      P1 & 59373.61016027& 17 &
      3138& 3666& $0.98\pm0.05$&
      $0.93\pm0.06$&  $1202.4\pm0.2$ & $12956^{143}_{137}$&$12298^{74}_{93}$&$30.0\pm3.1$&$-2.6\pm3.0$ \\
      P2 & 59373.61196044 & 21 & 2792&3736& $5.10\pm0.30$& $4.83\pm0.39$& $1209.6\pm0.2$& $12556^{58}_{54}$&$12523^{40}_{42}$&$19.4\pm1.1$&$-1.1\pm1.0$ \\
      P3 & 59373.65276972 & 19 & 2509 & 3117 & $1.34\pm0.05$& $1.73\pm0.07$ &$1211.4\pm0.5$& $11756^{31}_{42}$& $11743^{34}_{37}$& $25.5\pm3.2$&$3.5\pm3.1$ \\
      P4 & 59384.63337770& 18 &
      3185& 3785& $2.42\pm0.39$&
      $7.84\pm1.85$&
      $1212.4\pm0.2$&$8054^{179}_{166}$&$8044^{19}_{25}$&$33.7\pm4.4$&$-9.2\pm4.2$\\
      P5&59400.43483313& 34 &
      2887&3439&$2.36\pm0.06$&
      $1.33\pm0.04$&$1205.7\pm0.2$&$10135^{76}_{102}$&$9608^{91}_{73}$&$15.0\pm1.8$&$3.1\pm1.8$ \\
      P6&59400.47865632& 32 &
      3413&4189&$4.18\pm0.43$&
      $2.73\pm0.05$ &$1206.4\pm0.2$&$9715^{115}_{88}$&$9908^{96}_{86}$&$24.4\pm2.2$&$-0.5\pm2.1$ \\
      P7&59588.83444570& 16 &
       2888&3560&$0.65\pm0.03$&
      $0.60\pm0.19$&$1186.0\pm0.2$&$-15518^{84}_{164}$&$-16081^{18}_{18}$&$56.7\pm12.5$&$1.7\pm10.8$\\
      P8&59588.90674632& 18 &
      3518&3974&$0.85\pm0.07$&
      $1.73\pm0.07$&$1186.4\pm0.3$&$-16358^{298}_{108}$&$-16289^{17}_{18}$&$53.2\pm14.0$&$-12.8\pm11.7$ \\
      
    \hline     

\label{mergedtable}
\end{longtable}

\end{landscape}
\begin{footnotesize}
\setlength{\tabcolsep}{5pt}
\begin{longtable}{cccc|cccc}
  \rmfamily\\
  \caption{\textbf{ DM measurements and polarisation limits          of 113 Parkes and 20 GBT bursts.}\\
           MJDs are in barycentric dynamical time (TDB),\\
           $DM_{stru}$ is the structure-maximizing DM reported by \texttt{DM\_PHASE} with 1$\sigma$ uncertainities.\\
           $L/I$ is the 3$\sigma$ upper limit of linear polarization fraction,\\
           $|V/I|$ is the 3$\sigma$ upper limit of absolute circular polarization.\label{tab:pks_dm}}\\
\hline
    MJD         & $DM_{stru}$        &      $L/I$  &   $|V/I|$  & MJD          & $DM_{stru}$             &  $L/I$  &  $|V/I|$  \\[1pt]
    Barycentric & (cm$^{-3}$ pc) &     $\%$  &    $\%$     & Barycentric  & (cm$^{-3}$ pc) &   $\%$     & $\%$  \\
\hline\\
\hline
59351.47501700 & 1218.5$\pm$0.6 & 0.17 & 0.04  &  59373.64420215 & 1204.9$\pm$0.2 & 0.23 & 0.01    \\
59351.50214313 & 1216.7$\pm$0.3 & 0.12 & 0.01   &  59373.64487000 & 1218.6$\pm$0.5 & 0.12 & 0.09    \\
59351.51107259 & 1214.5$\pm$0.5 & 0.50 & 0.10  &  59373.64569171 & 1206.9$\pm$0.4 & 0.16 & 0.00    \\
59351.51688799 & 1214.2$\pm$0.3 & 0.35 & 0.14   &  59373.64569211 & 1204.1$\pm$0.4 & 0.16 & 0.10    \\
59351.53159300 & 1190.7$\pm$0.6 & 0.16 & 0.12  &  59373.65219555 & 1210.4$\pm$0.6 & 0.13 & 0.01    \\
59360.49552295 & 1207.6$\pm$0.8 & 0.12 & 0.04   &  59373.65263631 & 1208.2$\pm$0.3 & 0.70 & 0.10    \\
59373.58387171 & 1195.0$\pm$0.4 & 0.27 & 0.13   &  59373.65276972 & 1211.4$\pm$0.5 & 0.00 & 0.00    \\
59373.58506840 & 1217.5$\pm$0.5 & 0.17 & 0.11   &  59373.65757784 & 1205.5$\pm$0.4 & 0.48 & 0.09    \\
59373.58527470 & 1210.2$\pm$0.4 & 0.28 & 0.26   &  59373.65892470 & 1221.3$\pm$0.4 & 0.18 & 0.09    \\
59373.58725056 & 1226.5$\pm$0.6 & 0.44 & 0.24   &  59373.66090389 & 1203.4$\pm$0.4 & 0.15 & 0.02    \\
59373.58775793 & 1201.3$\pm$0.4 & 0.12 & 0.04   &  59373.66136040 & 1200.6$\pm$0.4 & 0.45 & 0.31    \\
59373.58953052 & 1205.3$\pm$0.3 & 0.43 & 0.06   &  59373.66216570 & 1225.8$\pm$0.3 & 0.31 & 0.08    \\
59373.58979583 & 1209.6$\pm$0.2 & 0.47 & 0.03   &  59373.66338565 & 1205.9$\pm$0.4 & 0.35 & 0.18    \\
59373.59164439 & 1193.6$\pm$0.3 & 0.20 & 0.23  &  59373.66531324 & 1229.5$\pm$0.4 & 0.44 & 0.17    \\
59373.59336195 & 1219.0$\pm$0.4 & 0.35 & 0.23  &  59373.66600046 & 1209.8$\pm$0.3 & 0.54 & 0.22    \\
59373.59346079 & 1230.9$\pm$0.6 & 0.16 & 0.22  &  59384.59149044 & 1191.4$\pm$0.6 & 0.45 & 0.33    \\
59373.59395317 & 1208.8$\pm$0.4 & 0.35 & 0.01  &  59384.60291739 & 1203.0$\pm$0.4 & 0.20 & 0.07    \\
59373.59568780 & 1222.5$\pm$0.7 & 0.36 & 0.24  &  59384.61696213 & 1219.2$\pm$0.4 & 0.35 & 0.17    \\
59373.59795912 & 1206.0$\pm$0.4 & 0.17 & 0.02  &  59384.62945819 & 1220.9$\pm$0.4 & 0.18 & 0.14    \\
59373.59813109 & 1201.7$\pm$0.4 & 0.27 & 0.20  &  59384.63064809 & 1204.4$\pm$0.5 & 0.40 & 0.33    \\
59373.60206856 & 1231.1$\pm$0.5 & 0.16 & 0.03  &  59384.63337770 & 1212.4$\pm$0.2 & 0.00 & 0.00    \\
59373.60244511 & 1229.6$\pm$0.4 & 0.22 & 0.09   &  59384.64653307 & 1218.3$\pm$0.4 & 0.31 & 0.06    \\
59373.60288415 & 1202.5$\pm$0.3 & 0.43 & 0.32   &  59384.65003630 & 1193.5$\pm$0.2 & 0.29 & 0.01    \\
59373.60528841 & 1204.0$\pm$0.3 & 0.10 & 0.00   &  59384.66461760 & 1203.0$\pm$0.6 & 0.34 & 0.11    \\
59373.60726872 & 1224.4$\pm$0.4 & 0.14 & 0.21  &  59400.42254010 & 1202.3$\pm$0.4 & 0.27 & 0.19    \\
59373.60726900 & 1224.4$\pm$0.4 & 0.16 & 0.15  &  59400.42940844 & 1206.6$\pm$0.3 & 0.25 & 0.04    \\
59373.60764212 & 1208.0$\pm$0.5 & 0.15 & 0.09   &  59400.43372431 & 1209.2$\pm$0.3 & 0.56 & 0.41    \\
59373.60920772 & 1196.6$\pm$0.4 & 0.38 & 0.19   &  59400.43483313 & 1205.7$\pm$0.2 & 0.00 & 0.00    \\
59373.60970619 & 1203.9$\pm$0.4 & 0.18 & 0.09  &  59400.44081726 & 1209.0$\pm$0.3 & 0.18 & 0.03    \\
59373.60991899 & 1210.9$\pm$0.6 & 0.45 & 0.08  &  59400.47363738 & 1211.0$\pm$0.2 & 0.10 & 0.07    \\
59373.61016027 & 1202.4$\pm$0.2 & 0.00 & 0.00   &  59400.47368864 & 1215.3$\pm$0.6 & 0.31 & 0.12    \\
59373.61196044 & 1209.6$\pm$0.2 & 0.00 & 0.00   &  59400.47865632 & 1206.4$\pm$0.2 & 0.00 & 0.00    \\
59373.61325821 & 1205.6$\pm$0.3 & 0.16 & 0.01  &  59453.20091796 & 1213.8$\pm$0.4 & 0.37 & 0.02    \\
59373.61346204 & 1224.7$\pm$0.4 & 0.57 & 0.07   &  59481.27941627 & 1193.4$\pm$0.6 & 0.22 & 0.08    \\
59373.61507133 & 1196.8$\pm$0.2 & 0.11 & 0.02  &  59481.31653004 & 1199.8$\pm$0.4 & 0.43 & 0.19    \\
59373.61581023 & 1208.9$\pm$0.5 & 0.11 & 0.09   &  59481.33072906 & 1184.8$\pm$0.3 & 0.62 & 0.07    \\
59373.61698667 & 1215.5$\pm$0.7 & 0.33 & 0.09  &  59562.99075975 & 1195.9$\pm$0.3 & 0.22 & 0.14    \\
59373.61867194 & 1211.0$\pm$0.3 & 0.25 & 0.22   &  59562.99127888 & 1195.3$\pm$0.7 & 0.25 & 0.17    \\
59373.61910203 & 1204.6$\pm$0.3 & 0.16 & 0.10   &  59562.99994392 & 1195.6$\pm$0.2 & 0.07 & 0.06    \\
59373.61928613 & 1217.2$\pm$0.2 & 0.15 & 0.20   &  59574.97810893 & 1200.9$\pm$0.2 & 0.25 & 0.02    \\
59373.61975100 & 1202.8$\pm$0.3 & 0.17 & 0.19  &  59574.98154916 & 1186.5$\pm$0.3 & 0.16 & 0.06    \\
59373.62016334 & 1209.7$\pm$0.3 & 0.27 & 0.08  &  59574.98319806 & 1191.2$\pm$0.3 & 0.13 & 0.04    \\
59373.62142998 & 1205.8$\pm$0.4 & 0.20 & 0.12   &  59574.99457007 & 1186.8$\pm$0.2 & 0.07 & 0.03    \\
59373.62189764 & 1212.7$\pm$0.2 & 0.19 & 0.04  &  59575.03001776 & 1193.7$\pm$0.3 & 0.29 & 0.13    \\
59373.62291764 & 1214.3$\pm$0.4 & 0.25 & 0.12   &  59575.03543462 & 1191.4$\pm$0.5 & 0.22 & 0.18    \\
59373.62368835 & 1190.4$\pm$0.4 & 0.25 & 0.23   &  59588.83444570 & 1186.0$\pm$0.2 & 0.00 & 0.00    \\
59373.62650076 & 1217.8$\pm$0.6 & 0.31 & 0.10  &  59588.85249719 & 1181.6$\pm$0.2 & 0.27 & 0.08    \\
59373.63150253 & 1215.2$\pm$0.3 & 0.39 & 0.26   &  59588.85839107 & 1190.6$\pm$0.3 & 0.13 & 0.14    \\
59373.63457117 & 1220.0$\pm$0.3 & 0.32 & 0.28   &  59588.86650268 & 1184.2$\pm$0.3 & 0.30 & 0.09    \\
59373.63659233 & 1228.2$\pm$0.4 & 0.27 & 0.19  &  59588.87421910 & 1185.0$\pm$0.4 & 0.14 & 0.01    \\
59373.63824900 & 1229.0$\pm$0.5 & 0.19 & 0.08  &  59588.89471775 & 1176.5$\pm$0.2 & 0.29 & 0.04    \\
59373.63880439 & 1201.6$\pm$0.6 & 0.06 & 0.04   &  59588.90358107 & 1201.9$\pm$0.3 & 0.34 & 0.04    \\
59373.64234183 & 1214.8$\pm$0.3 & 0.29 & 0.14  &  59588.90406449 & 1183.0$\pm$0.5 & 0.28 & 0.02    \\
59373.64276972 & 1200.4$\pm$0.4 & 0.31 & 0.06   &  59588.90674632 & 1186.4$\pm$0.2 & 0.00 & 0.00    \\
59373.64390840 & 1201.8$\pm$0.5 & 0.16 & 0.10   &  59588.91176179 & 1197.5$\pm$0.5 & 0.53 & 0.42    \\
59373.64406409 & 1212.6$\pm$0.4 & 0.16 & 0.08   &  59588.92442799 & 1188.5$\pm$0.5 & 0.34 & 0.02    \\
59373.64420209 & 1204.9$\pm$0.2 & 0.32 & 0.04   &     \\
\hline\\
\\
\hline
  MJD         & $DM_{stru}$             &  $L/I$  &   $|V/I|$  & MJD          & $D_{stru}$             &  $L/I$  & $|V/I|$  \\[1pt]
  Barycentric & (cm$^{-3}$ pc) &   $\%$    &   $\%$     & Barycentric & (cm$^{-3}$ pc) &   $\%$    &  $\%$\\ 
\hline
59106.86907048   &   1184.0$\pm$0.5 & 0.4 &0.16  &   59109.94361260	 & 1202.9$\pm$0.4 & 0.90 & 0.20     \\
59106.89390624	 & 1195.3$\pm$0.2 & 0.22 & 0.06   &  59109.98434067 & 1205.0$\pm$0.2 & 0.11 & 0.05  \\
59106.90344833	 & 1205.3$\pm$0.4 & 0.21 & 0.11  &   59110.04206052 & 1200.7$\pm$0.4 & 0.30 & 0.14\\
59106.90577597     & 1238.2$\pm$0.5 & 0.16 & 0.03  & 59110.04376341 & 1185.6$\pm$0.8 & 0.50 & 0.40    \\
59106.90960819	 & 1259.3$\pm$0.3 & 0.40 & 0.10  &   59522.69901763 & 1209.6$\pm$0.0 & 0.20 & 0.19  \\
59109.85579395	 & 1197.0$\pm$0.6 & 0.60 & 0.30   &  59522.70600676 & 1170.9$\pm$0.3 & 0.30 & 0.30   \\
59109.8633426     & 1228.2$\pm$0.8 & 0.60 & 0.50   & 59522.71448526 & 1208.5$\pm$0.3 & 0.40 & 0.20  \\
59109.87577747	 & 1198.6$\pm$0.1 & 0.09 & 0.02   &  59522.71514831 & 1226.9$\pm$0.3 & 0.20 &0.11  \\
59109.88086212	& 1207.0$\pm$0.8 & 0.18 & 0.11  &    59522.78008166 & 1231.8$\pm$0.6 & 0.50 & 0.30   \\
59109.92907860   & 1201.1$\pm$0.5 & 0.19 & 0.14   &  59522.89241628 & 1213.4$\pm$0.6 & 0.40 & 0.20   \\
\hline
\label{DM_pol_table}
\end{longtable}
\end{footnotesize}

\clearpage

\clearpage

\clearpage

\end{document}